\documentclass[12pt]{article}

\usepackage{mathrsfs}
\usepackage[T1]{fontenc}
\usepackage{mathpazo}
\usepackage{setspace}
\usepackage{cite}
\usepackage{amsfonts}
\usepackage{amssymb}
\usepackage{amsmath}
\usepackage{enumitem}
\usepackage{esint}                   
\usepackage{epsfig}
\usepackage{latexsym}
\usepackage{color}
\usepackage{graphicx}
\usepackage{hyperref}
\usepackage{caption}
\usepackage{subcaption}
\usepackage{nicefrac}
\usepackage[latin1]{inputenc}
\usepackage{pstricks}
\usepackage{slashed}
\usepackage{multirow}

\def\hybrid{\topmargin -30pt    \oddsidemargin 0pt 
        \headheight 0pt \headsep 0pt
        \textwidth 6.25in       
        \textheight 9.5in       
        \marginparwidth .875in
        \parskip 5pt plus 1pt   \jot = 1.5ex}

\hybrid

\def\baselinestretch{1.2}

\catcode`\@=11

\def\marginnote#1{}
\newcount\hour
\newcount\minute
\newtoks\amorpm
\hour=\time\divide\hour by60
\minute=\time{\multiply\hour by60 \global\advance\minute by-\hour}
\edef\standardtime{{\ifnum\hour<12 \global\amorpm={am}%
        \else\global\amorpm={pm}\advance\hour by-12 \fi
        \ifnum\hour=0 \hour=12 \fi
        \number\hour:\ifnum\minute<10 0\fi\number\minute\the\amorpm}}
\edef\militarytime{\number\hour:\ifnum\minute<10 0\fi\number\minute}

\def\draftlabel#1{{\@bsphack\if@filesw {\let\thepage\relax
   \xdef\@gtempa{\write\@auxout{\string
      \newlabel{#1}{{\@currentlabel}{\thepage}}}}}\@gtempa
   \if@nobreak \ifvmode\nobreak\fi\fi\fi\@esphack}
        \gdef\@eqnlabel{#1}}
\def\@eqnlabel{}
\def\@vacuum{}
\def\draftmarginnote#1{\marginpar{\raggedright\scriptsize\tt#1}}

\def\draft{\oddsidemargin -.5truein
        \def\@oddfoot{\sl preliminary draft \hfil
        \rm\thepage\hfil\sl\today\quad\militarytime}
        \let\@evenfoot\@oddfoot \overfullrule 3pt
        \let\label=\draftlabel
        \let\marginnote=\draftmarginnote
   \def\@eqnnum{(\theequation)\rlap{\kern\marginparsep\tt\@eqnlabel}%
\global\let\@eqnlabel\@vacuum}  }

\def\draft2{
        \def\@oddfoot{\sl preliminary draft \hfil
        \rm\thepage\hfil\sl\today\quad\militarytime}
        \let\@evenfoot\@oddfoot \overfullrule 3pt
        \let\label=\draftlabel
        \let\marginnote=\draftmarginnote
   \def\@eqnnum{(\theequation)\rlap{\kern\marginparsep\tt\@eqnlabel}%
\global\let\@eqnlabel\@vacuum}  }

\def\preprint{\twocolumn\sloppy\flushbottom\parindent 2em
        \leftmargini 2em\leftmarginv .5em\leftmarginvi .5em
        \oddsidemargin -.5in    \evensidemargin -.5in
        \columnsep .4in \footheight 0pt
        \textwidth 10.in        \topmargin  -.4in
        \headheight 12pt \topskip .4in
        \textheight 6.9in \footskip 0pt
        \def\@oddhead{\thepage\hfil\addtocounter{page}{1}\thepage}
        \let\@evenhead\@oddhead \def\@oddfoot{} \def\@evenfoot{} }

\def\numberbysection{\@addtoreset{equation}{section}
        \def\theequation{\thesection.\arabic{equation}}}

\def\underline#1{\relax\ifmmode\@@underline#1\else
        $\@@underline{\hbox{#1}}$\relax\fi}

\def\titlepage{\@restonecolfalse\if@twocolumn\@restonecoltrue\onecolumn
     \else \newpage \fi \thispagestyle{empty}\c@page\z@
        \def\thefootnote{\fnsymbol{footnote}} }

\def\endtitlepage{\if@restonecol\twocolumn \else \newpage \fi
        \def\thefootnote{\arabic{footnote}}
        \setcounter{footnote}{0}}  

\catcode`@=12
\relax

\def\figcap{\section*{Figure Captions\markboth
        {FIGURECAPTIONS}{FIGURECAPTIONS}}\list
        {Figure \arabic{enumi}:\hfill}{\settowidth\labelwidth{Figure
999:}
        \leftmargin\labelwidth
        \advance\leftmargin\labelsep\usecounter{enumi}}}
 \relax
\def\tablecap{\section*{Table Captions\markboth
        {TABLECAPTIONS}{TABLECAPTIONS}}\list
        {Table \arabic{enumi}:\hfill}{\settowidth\labelwidth{Table
999:}
        \leftmargin\labelwidth
        \advance\leftmargin\labelsep\usecounter{enumi}}}
 \relax
\def\reflist{\section*{References\markboth
        {REFLIST}{REFLIST}}\list
        {[\arabic{enumi}]\hfill}{\settowidth\labelwidth{[999]}
        \leftmargin\labelwidth
        \advance\leftmargin\labelsep\usecounter{enumi}}}
 \relax
\makeatletter
\newcounter{pubctr}
\def\publist{\@ifnextchar[{\@publist}{\@@publist}}
\def\@publist[#1]{\list
        {[\arabic{pubctr}]\hfill}{\settowidth\labelwidth{[999]}
        \leftmargin\labelwidth
        \advance\leftmargin\labelsep
        \@nmbrlisttrue\def\@listctr{pubctr}
        \setcounter{pubctr}{#1}\addtocounter{pubctr}{-1}}}
\def\@@publist{\list
        {[\arabic{pubctr}]\hfill}{\settowidth\labelwidth{[999]}
        \leftmargin\labelwidth
        \advance\leftmargin\labelsep
        \@nmbrlisttrue\def\@listctr{pubctr}}}
 \relax
\makeatother

\def\be{\begin{equation}}
\def\ee{\end{equation}}
\def\ba{\begin{eqnarray}}
\def\ea{\end{eqnarray}}

\def\r{\rho}
\def\a{\alpha}

\def\b{\beta}

\def\g{\gamma}

\def\th{\theta}
\def\Th{\Theta}

\def\l{\lambda}

\def\s{\sigma}
\def\S{\Sigma}

\def\cE{{\cal E}}
\def\cF{{\cal F}}
\def\cG{{\cal G}}

\def\cN{{\cal N}}

\def\cT{{\cal T}}
\def\cV{{\cal V}}

\def\no{\noindent}

\def\IR{\relax{\rm I\kern-.18em R}}

\def\inv{^{\raise.0ex\hbox{${\scriptscriptstyle -}$}\kern-.05em 1}}

\begin{document}


\renewcommand{\theequation}{\thesection.\arabic{equation}}
\csname @addtoreset\endcsname{equation}{section}

\begin{titlepage}
\begin{center}

\hfill HU-EP-26/09

\phantom{xx}
\vskip 0.5in

{\large \bf Holographic observables in TsT deformations of confining theories}

\vskip 0.5in

{\bf Madison Hammond}${}^{1a}$ \phantom{x}   and \phantom{x} {\bf Georgios Itsios}${}^{2b}$ \vskip 0.1in

${}^1$ Centre for Quantum Fields and Gravity, Department of Physics, Swansea University, Swansea, SA2 8PP, Wales, United Kingdom\\

\vskip .2in

${}^2$ Institut f\"{u}r Physik, Humboldt-Universit\"{a}t zu Berlin,\\
IRIS Geb\"{a}ude, Zum Gro{\ss}en Windkanal 2, 12489 Berlin, Germany

\end{center}

\vskip .4in

\centerline{\bf Abstract}

\no
We construct new families of type-IIB supergravity solutions by employing TsT transformations on the ten-dimensional geometry that arises after the uplift of the five-dimensional soliton solution of Anabal\'on, Nastase, and Oyarzo. In particular, we identify two marginal and two dipole deformations of the uplifted geometry. We then analyse a plethora of holographic observables---including Wilson loops, `t~Hooft loops, Page charges, entanglement entropy, and central charge---and compare their behaviour across the different deformed backgrounds.

\vfill
\no
 {
$^a$m.hammond.2412736@swansea.ac.uk,\\
$^b$georgios.itsios@physik.hu-berlin.de
}

\end{titlepage}
\vfill
\eject



\def\baselinestretch{1.2}
\baselineskip 20 pt

\newcommand{\eqn}[1]{(\ref{#1})}

\tableofcontents

\section{Introduction}

The gauge/gravity correspondence \cite{Maldacena:1997re} remains the main framework for studying strongly coupled quantum field theories, which are associated with type-II supergravity backgrounds.
In particular, solutions obtained from consistent truncations of type-IIB supergravity on deformed five-spheres have played a central r\^ole in understanding holographic RG flows, Coulomb-branch dynamics, and marginal deformations of $\mathcal{N}=4$ SYM \cite{Girardello:1999hj,Polchinski:2000uf,Klebanov:2000hb}.

There are two main ways in which the duality has been generalised: one involving wrapped branes on internal cycles \cite{Maldacena:2000yy,Atiyah:2000zz,Edelstein:2001pu,Nunez:2001pt,Gauntlett:2001ps,Bigazzi:2001aj,Klebanov:1998hh}, and the other involving D-branes placed at the tip of a conifold \cite{Klebanov:1998hh,Klebanov:2000nc,Klebanov:2000hb,Gubser:2004qj,Butti:2004pk}. In many such holographic constructions, the ultraviolet behaviour of the dual theory is controlled by higher-dimensional dynamics that does not correspond to a well-defined quantum field theory in the UV, necessitating a full string-theoretic completion. Gravitational systems capturing these features often arise as extensions of the original duality between type-II string theory on $AdS_5 \times S^5$ and $\mathcal{N} = 4$ Supersymmetric Yang-Mills (SYM) theory on $\mathbb{R}^{1,3}$. A particularly useful example is provided by generalisations of the AdS soliton of \cite{Anabalon:2021tua}, where the geometry terminates smoothly at a finite value of the radial coordinate, generating a mass gap in the dual field theory. This background corresponds to $\mathcal{N} = 4$ SYM compactified on $\mathbb{R}^{1,2} \times S^1$.

The AdS soliton is a smooth, cigar-like geometry in the infrared obtained via a double Wick rotation of a black hole solution. This has been studied vastly in \cite{Chatzis:2024kdu,Chatzis:2024top,macpherson2024,Fatemiabhari:2024aua,Chatzis:2025dnu,Chatzis:2025hek,Nunez:2025gxq,Nunez:2025puk,Macpherson:2024frt,Kumar:2024pcz,Macpherson:2025pqi,Macpherson:2024qfi,Castellani:2024pmx,Castellani:2024ial,Chatzis:2025wfv,Whittle:2025yog}. The soliton of \cite{Anabalon:2024che} generalizes the supersymmetric soliton introduced in \cite{Anabalon:2021tua} by incorporating a non-trivial scalar profile, an additional gauge field, and an extra parameter. This wider class of solutions has been interpreted as describing Coulomb-branch deformations of $\mathcal{N} = 4$ SYM \cite{Freedman:1999gk,Gubser:2000nd}. The corresponding five-dimensional gauged supergravity soliton and its uplift to type-IIB supergravity, further explored in \cite{Chatzis:2025dnu,Chatzis:2025hek}, is used as the seed for constructing new holographic duals via the TsT solution-generating technique \cite{Lunin:2005jy}. This seed background describes supersymmetry-preserving RG flows from a four-dimensional superconformal field theory (SCFT) in the UV to a three-dimensional quantum field theory in the IR. It exhibits confinement of external quarks and possesses a mass gap. 

In this work we apply TsT transformations to the uplifted soliton of \cite{Anabalon:2024che}, taking place along several distinct pairs of the available $U(1)$ isometry directions. As a result we find various ten-dimensional geometries that are dual to marginal or dipole deformations of the quantum field theory associated to the seed type-IIB background.

As shown in \cite{Chatzis:2025dnu,Chatzis:2025hek}, many holographic observables in this class of backgrounds display a universal factorisation: one component depends only on the UV SCFT data, while the other encodes the RG-flow dynamics and is insensitive to the choice of UV fixed point. We find that the marginally deformed backgrounds retain several of these universal features. 

The remainder of the paper is organised as follows. In Section 2 we review the uplift of the 5d soliton found in \cite{Anabalon:2024che} to type-IIB supergravity, and construct its marginal deformations. Section 3 analyses the holographic observables and compares their behaviour across the various backgrounds. We conclude and summarise in Section 4.

\section{Supergravity backgrounds}
\label{SupergravityBackgrounds}

In this section, we review the content of a type-IIB supergravity background, obtained by uplifting a solution of five-dimensional gauged supergravity involving three gauge fields and two scalar fields \cite{Anabalon:2024che}. The five-dimensional geometry is supplemented with an internal space, which represents a deformation of a five-dimensional sphere through the gauge fields. This background has several manifest $U(1)$ isometries, three of which lie in the deformed sphere and the rest in the external space. The presence of these isometries allows for the construction of new type-IIB solutions via TsT transformations. Such transformations require the existence of at least two $U(1)$ isometries in the geometry, corresponding to symmetries under shifts in the associated coordinates (denoted here as $\Theta_1$ and $\Theta_2$), and can be summarised in the following three steps
\begin{itemize}
 \item \textbf{Step 1:} Perform a T-duality along $\Th_1$.
 
 \item \textbf{Step 2:} Implement the coordinate shift $\Th_2 \to \Th_2 + \g \, \Th_1$, where $\g$ is a real deformation parameter.
 
 \item \textbf{Step 3:} Perform a second T-duality along $\Th_1$.
\end{itemize}
Depending on whether the $U(1)$ directions lie in the internal or external space, the outcome of the TsT transformation is understood as the gravity dual of a marginal, dipole, or non-commutative deformation of a QFT. Below, we start from the solution of \cite{Anabalon:2024che} and then proceed to the geometries generated via TsT operations, focusing on those corresponding to marginal and dipole deformations.

\subsection{The original background}
\label{solutionANO}

The geometry of the type-IIB background found in \cite{Anabalon:2024che} and further studied in \cite{Chatzis:2025dnu,Chatzis:2025hek} is only supported by a self-dual Ramond-Ramond (RR) five-form, while the other fields are trivial. The ten-dimensional metric takes the form
\footnote{
In this parametrisation, the periodicities of the angles of the deformed five-sphere are
\[
 \th, \psi\in \Big[ 0 , \frac{\pi}{2} \Big] \, , \qquad \phi_i \in [0 , 2 \pi] \quad (i = 1,2,3) \, .
\]
}
\begin{equation}
 \label{metricANO}
 \begin{aligned}
  ds^2 & = \frac{\zeta}{L^2} \Big( r^2 \big( - dt^2 + dw^2 + dz^2 + L^2 F \, d\phi^2 \big) + L^2 \frac{dr^2}{r^2 F \l^6} + L^4 d\th^2 \Big)
  \\[5pt]
   & + \frac{L^2}{\zeta} \left( \cos^2\th \, d\psi^2 + \cos^2\th \, \sin^2\psi \Big( d\phi_1 + \frac{A_1}{L} \Big)^2 + \cos^2\th \, \cos^2\psi \Big( d\phi_2 + \frac{A_2}{L} \Big)^2 \right.
   \\[5pt]
   & \left. + \l^6 \sin^2\th \Big( d\phi_3 + \frac{A_3}{L} \Big)^2 \right) \, ,
 \end{aligned}
\end{equation}
where the functions $\lambda$ and $F$ depend exclusively on the radial coordinate $r$, while $\zeta$ is a function of both $r$ and the angular coordinate $\theta$. These functions are explicitly given by
\begin{equation}
 \begin{aligned}
  & \zeta(r, \th) = \sqrt{1 + \varepsilon \frac{\ell^2}{r^2} \cos^2\th} \, , \qquad
      \l(r) = \Big( 1 + \varepsilon \frac{\ell^2}{r^2} \Big)^{\frac{1}{6}} \, ,
      \\[5pt]
  & F(r) = \frac{1}{L^2} - \varepsilon \frac{\ell^2 L^2}{r^4} \Big( q^2_1 - \frac{q^2_2}{\l^6} \Big) \, ,
 \end{aligned}
\end{equation}
where $\ell$, $q_1$, and $q_2$ are constant parameters, and $\varepsilon = \pm 1$ is a sign which determines two branches of the solution. The one-forms $A_i \, (i = 1, 2, 3)$ correspond to the gauge fields in the five-dimensional gauged supergravity theory, and are given by
\begin{equation}
 \label{gaugeFields}
 A_1 = A_2 = \varepsilon q_1 \ell^2 L \frac{r^2 - r^2_{\star}}{r^2 r^2_{\star}} \, d\phi \, , \qquad 
 A_3 = \varepsilon q_2 \ell^2 L \frac{r^2 - r^2_{\star}}{\big( r^2 + \varepsilon \ell^2 \big) \big( r^2_{\star} + \varepsilon \ell^2 \big)} \, d\phi \, .
\end{equation}
Here, $r_{\star}$ stands for the largest root of $F(r)$ and corresponds to the end of the space, meaning that $r \geq r_\star$.

In order to better present the RR five-form we introduce the following orthogonal frame
\footnote{
In this frame, the line element \eqref{metricANO} reads $ds^2 = \eta_{ab} e^a e^b$, where $\eta_{ab}$ is the ten-dimensional Minkowski metric.
}
\begin{equation}
 \label{frameANO}
 \begin{aligned}
   & e^0 = \frac{\sqrt{\zeta}}{L} \, r \, dt \, , \qquad
      e^1 = \frac{\sqrt{\zeta}}{L} \, r \, dw \, , \qquad 
      e^2 = \frac{\sqrt{\zeta}}{L} \, r \, dz \, , \qquad
      e^3 = \sqrt{\zeta F} \, r \, d\phi \, ,
      \\[5pt]
   & e^4 = \sqrt{\frac{\zeta}{F}} \, \frac{dr}{\l^3 r} \, , \qquad
      e^5 = L \sqrt{\zeta} \, d\th \, , \qquad
      e^6 = L \frac{\cos\th}{\sqrt{\zeta}} \, d\psi \, ,
      \\[5pt]
   & e^7 = L \frac{\cos\th \sin\psi}{\sqrt{\zeta}} \, \Big( d\phi_1 + \frac{A_1}{L} \Big) \, , \qquad
       e^8 = L \frac{\cos\th \cos\psi}{\sqrt{\zeta}} \, \Big( d\phi_2 + \frac{A_2}{L} \Big) \, ,
       \\[5pt]
    & e^9 = L \frac{\l^3}{\sqrt{\zeta}} \sin\th \Big( d\phi_3 + \frac{A_3}{L} \Big) \, .
 \end{aligned}
\end{equation}
In this basis, the RR five-form $F_5$ becomes
\begin{equation}
 \label{F5ANO}
 F_5 = (1 + \star) G_5 \, ,
\end{equation}
with
\begin{equation}
 \label{G5ANO}
 \begin{aligned}
  G_5 & = \frac{2}{L} \frac{\l^3 (1 + \zeta^2)}{\zeta^{\frac{5}{2}}} \, e^{01234}
            - \varepsilon \ell^2 \frac{\sqrt{F}}{r^2 \zeta^{\frac{5}{2}}} \sin(2 \th) \, e^{01235}
            - \frac{A'_{1\phi}}{\zeta^{\frac{3}{2}}} \sin\th \sin\psi \, e^{01257}
            \\[5pt]
           & - \frac{A'_{1\phi}}{\zeta^{\frac{3}{2}}} \sin\th \cos\psi \, e^{01258}
              + \frac{A'_{1\phi}}{\sqrt{\zeta}} \cos\psi \, e^{01267}
               - \frac{A'_{1\phi}}{\sqrt{\zeta}} \sin\psi \, e^{01268}
               + \frac{\l^9 A'_{3\phi}}{\zeta^{\frac{3}{2}}} \cos\th \, e^{01259} \, .
 \end{aligned}
\end{equation}
In the above we adopt the notation $e^{a_1 \ldots a_5} = e^{a_1} \wedge \ldots \wedge e^{a_5}$. Moreover, $A'_{i\phi} \, (i = 1, 2, 3)$ refer to the derivatives of the gauge field $\phi$-components \eqref{gaugeFields} with respect to the radial coordinate $r$.

In the following we will focus on the supersymmetric case where $q_1 = q_2 = Q$. In this case, $Q$ is related to the root $r_\star$ of the function $F(r)$ as
\begin{equation}
    Q = \frac{r^3_\star}{L^2 \ell^2} \sqrt{1 + \hat{\nu}} \, , \qquad \hat{\nu} := \varepsilon \frac{\ell^2}{r^2_\star} \, .
\end{equation}
Notice that $Q$ is real as long as $\hat{\nu} \geq - 1$. This is always the case in the branch $\varepsilon = + 1$. However, when $\varepsilon = - 1$ this implies that $\ell \leq r_\star$. One can check that the geometry \eqref{metricANO} has a vanishing Ricci scalar. However, the curvature invariants $R_{\mu\nu} R^{\mu\nu}$ and $R_{\mu\nu\r\s} R^{\mu\nu\r\s}$ diverge along the curve
\begin{equation}
    \xi^2 + \hat{\nu} \cos^2\th = 0 \, , \qquad \xi := \frac{r}{r_\star} \, .
\end{equation}
Since $r \geq r_\star$, the geometry is singular only when $\varepsilon = - 1$. Taking also into account that $\hat{\nu} \geq - 1$, we find that the singularity is manifest when
\begin{equation}
    \xi = 1 \, , \qquad \th = 0 \, , \qquad \hat{\nu} = - 1 \, .
\end{equation}
This is equivalent to $r = r_\star = \ell$ at $\th = 0$,  with $\hat{\nu} = - 1$ and $Q = 0$
\footnote{Notice that the function $F$ can be written as
\[
 \begin{aligned}
     & F(\xi) = \frac{\big( \xi^2 - 1 \big) \Big( \xi^4 + \left( 1 + \hat{\nu} \right) \left( 1 + \xi^2 \right) \Big)}{L^2 \xi^4 \big( \xi^2 + \hat{\nu} \big)} = \frac{\big( \xi^2 - 1 \big) \big( \xi^2 - \xi^2_+ \big) \big( \xi^2 - \xi^2_- \big)}{L^2 \xi^4 \big( \xi^2 + \hat{\nu} \big)} \, ,
     \\[5pt]
& \xi^2_\pm = \frac{1}{2} \Big( - 1 - \hat{\nu} \pm \sqrt{\big( 1 + \hat{\nu} \big) \big( \hat{\nu} - 3 \big)} \Big) \, .
 \end{aligned}
\]
In the \emph{degenerate} case where $\hat{\nu} = - 1$ we have that $\xi_\pm = 0$ and $F(\xi) = \frac{1}{L^2}$.
}.

The periodicity of the confining direction $\phi$ is fixed by requiring the absence of conical singularities in the $\xi - \phi$ plane. One can check that near $\xi = 1$ the metric along the $(\xi , \phi)$ directions becomes
\begin{equation}
    ds^2 = \sqrt{1 + \hat{\nu} \cos^2\th} \left( d\r^2 + \frac{r^2_\star}{L^4} \frac{\big( 3 + 2 \hat{\nu} \big)^2}{1 + \hat{\nu}} \r^2 d\phi^2 \right) \, , \qquad
    \r = L \sqrt{\frac{2}{3 + 2 \hat{\nu}} (\xi - 1)} \, .
\end{equation}
To avoid the conical singularity, one has to restrict $\phi$ in $[0 , L_{\phi}]$, where
\begin{equation}
\label{phiPeriodicity}
    L_{\phi} = 2 \pi \frac{L^2}{r_\star} \frac{\sqrt{1 + \hat{\nu}}}{3 + 2 \hat{\nu}} \, .
\end{equation}
A more detailed discussion on the geometry can be found in \cite{Chatzis:2025dnu}.

The type-IIB solution described above exhibits six $U(1)$ isometries, which correspond to symmetries of the geometry under shifts in the angular directions $(\phi_1, \, \phi_2, \phi_3)$, as well as $z$, $w$ and the compactified coordinate $\phi$. Notice that the background remains invariant under the exchange $\phi_1 \leftrightarrow \phi_2$, provided that $\psi \to \frac{\pi}{2} - \psi$. This suggests that the coordinates $\phi_1$ and $\phi_2$ can be treated on equal footing in the various computations in which they are involved. In the following we exploit the presence of these isometries to generate new solutions of the type-IIB supergravity via TsT deformations.

\subsection{Marginal deformation I}
\label{solutionTsT121}

Let us now present a background that is found by applying two T-dualities along $\phi_1$ and a shift along $\phi_2$. The effect of this transformation in the geometry \eqref{metricANO} is that the frame components $e^7$ and $e^8$ in \eqref{frameANO} are mapped to
\begin{equation}
\label{frameDeformationTsT121}
 (e^7 , e^8) \mapsto \frac{1}{W} \, (e^7 , e^8) \, , \qquad W = \sqrt{1 + \g^2 L^4 \frac{\cos^4\th \, \cos^2\psi \, \sin^2\psi}{\zeta^2}} \, .
\end{equation}
In particular, the new frame is
\begin{equation}
 \label{frameTsT121}
 \begin{aligned}
   & e^0 = \frac{\sqrt{\zeta}}{L} \, r \, dt \, , \qquad
      e^1 = \frac{\sqrt{\zeta}}{L} \, r \, dw \, , \qquad 
      e^2 = \frac{\sqrt{\zeta}}{L} \, r \, dz \, , \qquad
      e^3 = \sqrt{\zeta F} \, r \, d\phi \, ,
      \\[5pt]
   & e^4 = \sqrt{\frac{\zeta}{F}} \, \frac{dr}{\l^3 r} \, , \qquad
      e^5 = L \sqrt{\zeta} \, d\th \, , \qquad
      e^6 = L \frac{\cos\th}{\sqrt{\zeta}} \, d\psi \, ,
      \\[5pt]
   & e^7 = L \frac{\cos\th \sin\psi}{W \, \sqrt{\zeta}} \, \Big( d\phi_1 + \frac{A_1}{L} \Big) \, , \qquad
       e^8 = L \frac{\cos\th \cos\psi}{W \, \sqrt{\zeta}} \, \Big( d\phi_2 + \frac{A_2}{L} \Big) \, ,
       \\[5pt]
    & e^9 = L \frac{\l^3}{\sqrt{\zeta}} \sin\th \Big( d\phi_3 + \frac{A_3}{L} \Big) \, .
 \end{aligned}
\end{equation}
Therefore, the deformed line element reads
\begin{equation}
 \label{metricTsT121}
 \begin{aligned}
  ds^2 & = \frac{\zeta}{L^2} \Big( r^2 \big( - dt^2 + dw^2 + dz^2 + L^2 F \, d\phi^2 \big) + L^2 \frac{dr^2}{r^2 F \l^6} + L^4 d\th^2 \Big)
  \\[5pt]
   & + \frac{L^2}{\zeta} \left( \cos^2\th \, d\psi^2 + \frac{\cos^2\th \, \sin^2\psi}{W^2} \Big( d\phi_1 + \frac{A_1}{L} \Big)^2 + \frac{\cos^2\th \, \cos^2\psi}{W^2} \Big( d\phi_2 + \frac{A_2}{L} \Big)^2 \right.
   \\[5pt]
   & \left. + \l^6 \sin^2\th \Big( d\phi_3 + \frac{A_3}{L} \Big)^2 \right) \, .
 \end{aligned}
\end{equation}
The Neveu-Schwarz (NS) sector of the deformed solution also contains a non-trivial dilaton
\begin{equation}
 \label{dilatonTsT121}
 \Phi = - \ln W \, ,
\end{equation}
and the non-trivial two-form
\begin{equation}
 B_2 = - \sqrt{W^2 - 1} \, e^7 \wedge e^8 \, .
\end{equation}
In the absence of deformation, i.e, when $\g = 0$, the dilaton and $B_2$ field vanish.

Moving to the RR sector, we find that the deformation generates a three-form which is
%
%
\begin{equation}
 \label{F3TsT121}
  F_3 = \sqrt{W^2 - 1} \left( - \varepsilon \ell^2 \frac{\sqrt{F}}{r^2 \zeta^{\frac{5}{2}}} \sin(2 \th) \, e^{469}
  - \frac{2}{L} \frac{\l^3 (1 + \zeta^2)}{\zeta^{\frac{5}{2}}} \, e^{569}
  + \frac{\l^9 A'_{3\phi}}{\zeta^{\frac{3}{2}}} \cos\th \, e^{346} \right) \, .
\end{equation}
In addition, there exists a self-dual five-form $F_5$ written as in \eqref{F5ANO}, where now
\begin{equation}
 \begin{aligned}
  G_5 & = \frac{2}{L} \frac{\l^3 (1 + \zeta^2)}{\zeta^{\frac{5}{2}}} \, e^{01234}
            - \varepsilon \ell^2 \frac{\sqrt{F}}{r^2 \zeta^{\frac{5}{2}}} \sin(2 \th) \, e^{01235}
            - \frac{W A'_{1\phi}}{\zeta^{\frac{3}{2}}} \sin\th \sin\psi \, e^{01257}
            \\[5pt]
           & - \frac{W A'_{1\phi}}{\zeta^{\frac{3}{2}}} \sin\th \cos\psi \, e^{01258}
              + \frac{W A'_{1\phi}}{\sqrt{\zeta}} \cos\psi \, e^{01267}
               - \frac{W A'_{1\phi}}{\sqrt{\zeta}} \sin\psi \, e^{01268}
             \\[5pt]
            & + \frac{\l^9 A'_{3\phi}}{\zeta^{\frac{3}{2}}} \cos\th \, e^{01259} \, .
 \end{aligned}
\end{equation}
Setting $\g = 0$, only the self-dual form survives. 

The equations of motion have been checked using Mathematica.

\subsection{Marginal deformation II}
\label{solutionTsT131}

A second possibility is to consider a deformation of the solution presented in Sec. \ref{solutionANO}, obtained by two T-dualities in $\phi_1$ and a coordinate shift in $\phi_3$
\footnote{
A deformation generated by a TsT transformation in the directions $\phi_2$ and $\phi_3$ is expected to be equivalent to the type-IIB solution of the present section, due to the symmetry of the seed background under the exchange $\phi_1 \leftrightarrow \phi_2$ (with $\psi \to \frac{\pi}{2} - \psi$).
}
. In this case, the frame components $e^7$ and $e^9$ in \eqref{frameANO} are mapped to
\begin{equation}
 \label{frameDeformationTsT131}
 (e^7 , e^9) \mapsto \frac{1}{W} \, (e^7 , e^9) \, , \qquad W = \sqrt{1 + \g^2 L^4 \frac{\l^6 \cos^2\th \, \sin^2\th \, \sin^2\psi}{\zeta^2}} \, ,
\end{equation}
while the rest are not affected. More precisely
\begin{equation}
 \label{frameTsT131}
 \begin{aligned}
   & e^0 = \frac{\sqrt{\zeta}}{L} \, r \, dt \, , \qquad
      e^1 = \frac{\sqrt{\zeta}}{L} \, r \, dw \, , \qquad 
      e^2 = \frac{\sqrt{\zeta}}{L} \, r \, dz \, , \qquad
      e^3 = \sqrt{\zeta F} \, r \, d\phi \, ,
      \\[5pt]
   & e^4 = \sqrt{\frac{\zeta}{F}} \, \frac{dr}{\l^3 r} \, , \qquad
      e^5 = L \sqrt{\zeta} \, d\th \, , \qquad
      e^6 = L \frac{\cos\th}{\sqrt{\zeta}} \, d\psi \, ,
      \\[5pt]
   & e^7 = L \frac{\cos\th \sin\psi}{W \, \sqrt{\zeta}} \, \Big( d\phi_1 + \frac{A_1}{L} \Big) \, , \qquad
       e^8 = L \frac{\cos\th \cos\psi}{\sqrt{\zeta}} \, \Big( d\phi_2 + \frac{A_2}{L} \Big) \, ,
       \\[5pt]
    & e^9 = L \frac{\l^3}{W \, \sqrt{\zeta}} \sin\th \Big( d\phi_3 + \frac{A_3}{L} \Big) \, .
 \end{aligned}
\end{equation}
As a result, the line element of the deformed geometry is
\begin{equation}
 \label{metricTsT131}
 \begin{aligned}
  ds^2 & = \frac{\zeta}{L^2} \Big( r^2 \big( - dt^2 + dw^2 + dz^2 + L^2 F \, d\phi^2 \big) + L^2 \frac{dr^2}{r^2 F \l^6} + L^4 d\th^2 \Big)
  \\[5pt]
   & + \frac{L^2}{\zeta} \left( \cos^2\th \, d\psi^2 + \frac{\cos^2\th \, \sin^2\psi}{W^2} \Big( d\phi_1 + \frac{A_1}{L} \Big)^2 + \cos^2\th \, \cos^2\psi \Big( d\phi_2 + \frac{A_2}{L} \Big)^2 \right.
   \\[5pt]
   & \left. + \frac{\l^6 \sin^2\th}{W^2} \Big( d\phi_3 + \frac{A_3}{L} \Big)^2 \right) \, .
 \end{aligned}
\end{equation}
The dilaton of this solution is again given by an expression of the form \eqref{dilatonTsT121}, where now the function $W$ is \eqref{frameDeformationTsT131}. The NS sector also contains a non-trivial two-form which reads
\begin{equation}
 B_2 = - \sqrt{W^2 - 1} \, e^7 \wedge e^9 \, .
\end{equation}
When $\g = 0$, the NS sector contains only the metric.

The RR sector consists of a three- and a five-form, with the latter being self-dual. For the three-form we find
\begin{equation}
 \label{F3TsT131}
 \begin{aligned}
     F_3 = \sqrt{W^2 - 1} & \left( \varepsilon \ell^2 \frac{\sqrt{F}}{r^2 \zeta^{\frac{5}{2}}} \sin(2 \th) \, e^{468}
     + \frac{2}{L} \frac{\l^3 (1 + \zeta^2)}{\zeta^{\frac{5}{2}}} \, e^{568}
     + \frac{A'_{1\phi}}{\zeta^{\frac{3}{2}}} \sin\th \cos\psi \, e^{346} \right.
     \\[5pt]
     & \left. - \frac{A'_{1\phi}}{\sqrt{\zeta}} \sin\psi \, e^{345} \right) \, .
 \end{aligned}
\end{equation}
The five-form can again be written as in \eqref{F5ANO}, where now
\begin{equation}
 \begin{aligned}
  G_5 & = \frac{2}{L} \frac{\l^3 (1 + \zeta^2)}{\zeta^{\frac{5}{2}}} \, e^{01234}
            - \varepsilon \ell^2 \frac{\sqrt{F}}{r^2 \zeta^{\frac{5}{2}}} \sin(2 \th) \, e^{01235}
            - \frac{W A'_{1\phi}}{\zeta^{\frac{3}{2}}} \sin\th \sin\psi \, e^{01257}
            \\[5pt]
           & - \frac{A'_{1\phi}}{\zeta^{\frac{3}{2}}} \sin\th \cos\psi \, e^{01258}
              + \frac{W A'_{1\phi}}{\sqrt{\zeta}} \cos\psi \, e^{01267}
               - \frac{A'_{1\phi}}{\sqrt{\zeta}} \sin\psi \, e^{01268}
            \\[5pt]
            & + \frac{W \l^9 A'_{3\phi}}{\zeta^{\frac{3}{2}}} \cos\th \, e^{01259} \, .
 \end{aligned}
\end{equation}
The equations of motion have been checked using Mathematica.

Obviously, when the deformation is not present the three-form is trivial.

\subsection{Dipole deformation I}
\label{solutionTsT1w1}

We now move to a type-IIB background generated via a TsT transformation along an internal and an external $U(1)$ direction, which is therefore dual to a dipole deformation of a QFT. In particular, we consider two T-dualities and a shift along the coordinates $\phi_1$ and $w$, respectively, applied to the solution of Sec. \ref{solutionANO}. Consequently, the TsT transformation maps the frame components $e^1$ and $e^7$ of eq. \eqref{frameANO} as follows
\begin{equation}
\label{frameDeformationTsT1w1}
 (e^1 , e^7) \mapsto \frac{1}{W} \, (e^1 , e^7) \, , \qquad W = \sqrt{1 + \g^2 r^2 \cos^2\th \sin^2\psi} \, .
\end{equation}
Therefore, the corresponding ten-dimensional frame reads
\begin{equation}
 \label{frameTsT1w1}
 \begin{aligned}
   & e^0 = \frac{\sqrt{\zeta}}{L} \, r \, dt \, , \qquad
      e^1 = \frac{\sqrt{\zeta}}{L \, W} \, r \, dw \, , \qquad 
      e^2 = \frac{\sqrt{\zeta}}{L} \, r \, dz \, , \qquad
      e^3 = \sqrt{\zeta F} \, r \, d\phi \, ,
      \\[5pt]
   & e^4 = \sqrt{\frac{\zeta}{F}} \, \frac{dr}{\l^3 r} \, , \qquad
      e^5 = L \sqrt{\zeta} \, d\th \, , \qquad
      e^6 = L \frac{\cos\th}{\sqrt{\zeta}} \, d\psi \, ,
      \\[5pt]
   & e^7 = L \frac{\cos\th \sin\psi}{W \, \sqrt{\zeta}} \, \Big( d\phi_1 + \frac{A_1}{L} \Big) \, , \qquad
       e^8 = L \frac{\cos\th \cos\psi}{\sqrt{\zeta}} \, \Big( d\phi_2 + \frac{A_2}{L} \Big) \, ,
       \\[5pt]
    & e^9 = L \frac{\l^3}{\sqrt{\zeta}} \sin\th \Big( d\phi_3 + \frac{A_3}{L} \Big) \, .
 \end{aligned}
\end{equation}
This results in the following deformed line element
\begin{equation}
 \label{metricTsT1w1}
 \begin{aligned}
  ds^2 & = \frac{\zeta}{L^2} \left( r^2 \left( - dt^2 + \frac{dw^2}{W^2} + dz^2 + L^2 F \, d\phi^2 \right) + L^2 \frac{dr^2}{r^2 F \l^6} + L^4 d\th^2 \right)
  \\[5pt]
   & + \frac{L^2}{\zeta} \left( \cos^2\th \, d\psi^2 + \frac{\cos^2\th \, \sin^2\psi}{W^2} \Big( d\phi_1 + \frac{A_1}{L} \Big)^2 + \cos^2\th \, \cos^2\psi \Big( d\phi_2 + \frac{A_2}{L} \Big)^2 \right.
   \\[5pt]
   & \left. + \l^6 \sin^2\th \Big( d\phi_3 + \frac{A_3}{L} \Big)^2 \right) \, .
 \end{aligned}
\end{equation}
In addition to the above geometry, the NS sector of the solution also contains a dilaton, which takes the usual form given in eq. \eqref{dilatonTsT121}, with $W$ as in Eq.~\eqref{frameDeformationTsT1w1}, and a two-form given by
\begin{equation}
 B_2 = \sqrt{W^2 - 1} \, e^1 \wedge e^7 \, .
\end{equation}
The latter vanishes upon setting $\g = 0$.

For the RR sector, we obtain the non-trivial three-form
\begin{equation}
 \label{F3TsT1w1}
  F_3 = \sqrt{W^2 - 1} \left(
  \frac{A'_{1\phi}}{\zeta^{\frac{3}{2}}} \sin\th \sin\psi \, e^{025}
  - \frac{A'_{1\phi}}{\sqrt{\zeta}} \cos\psi \, e^{026}
  \right) \, .
\end{equation}
As expected, the above three-form vanishes in the absence of deformation. The RR sector also contains a self-dual five-form, which can be written as in eq. \eqref{F5ANO}, with
\begin{equation}
 \begin{aligned}
  G_5 & = \frac{2}{L} \frac{W \l^3 (1 + \zeta^2)}{\zeta^{\frac{5}{2}}} \, e^{01234}
            - \varepsilon \ell^2 \frac{W\sqrt{F}}{r^2 \zeta^{\frac{5}{2}}} \sin(2 \th) \, e^{01235}
            - \frac{A'_{1\phi}}{\zeta^{\frac{3}{2}}} \sin\th \sin\psi \, e^{01257}
            \\[5pt]
           & - \frac{W A'_{1\phi}}{\zeta^{\frac{3}{2}}} \sin\th \cos\psi \, e^{01258}
              + \frac{A'_{1\phi}}{\sqrt{\zeta}} \cos\psi \, e^{01267}
               - \frac{W A'_{1\phi}}{\sqrt{\zeta}} \sin\psi \, e^{01268}
            \\[5pt]
            & + \frac{W \l^9 A'_{3\phi}}{\zeta^{\frac{3}{2}}} \cos\th \, e^{01259} \, .
 \end{aligned}
\end{equation}
When $\g = 0$, the above expression coincides with that of eq. \eqref{G5ANO}.

\subsection{Dipole deformation II}
\label{solutionTsT3w3}

We continue with a type-IIB background generated via a TsT transformation along the $U(1)$ directions $\phi_3$ and $w$, applied to the solution of Sec. \ref{solutionANO}. Such a transformation affects the frame components $e^1$ and $e^9$ of eq. \eqref{frameANO}, which now transform as
\begin{equation}
\label{frameDeformationTsT3w3}
 (e^1 , e^9) \mapsto \frac{1}{W} \, (e^1 , e^9) \, , \qquad
 W = \sqrt{1 + \g^2 \l^6 r^2 \sin^2\th} \, .
\end{equation}
The ten-dimensional frame of the deformed geometry becomes
\begin{equation}
 \label{frameTsT3w3}
 \begin{aligned}
   & e^0 = \frac{\sqrt{\zeta}}{L} \, r \, dt \, , \qquad
      e^1 = \frac{\sqrt{\zeta}}{L \, W} \, r \, dw \, , \qquad 
      e^2 = \frac{\sqrt{\zeta}}{L} \, r \, dz \, , \qquad
      e^3 = \sqrt{\zeta F} \, r \, d\phi \, ,
      \\[5pt]
   & e^4 = \sqrt{\frac{\zeta}{F}} \, \frac{dr}{\l^3 r} \, , \qquad
      e^5 = L \sqrt{\zeta} \, d\th \, , \qquad
      e^6 = L \frac{\cos\th}{\sqrt{\zeta}} \, d\psi \, ,
      \\[5pt]
   & e^7 = L \frac{\cos\th \sin\psi}{\sqrt{\zeta}} \, \Big( d\phi_1 + \frac{A_1}{L} \Big) \, , \qquad
       e^8 = L \frac{\cos\th \cos\psi}{\sqrt{\zeta}} \, \Big( d\phi_2 + \frac{A_2}{L} \Big) \, ,
       \\[5pt]
    & e^9 = L \frac{\l^3}{W \, \sqrt{\zeta}} \sin\th \Big( d\phi_3 + \frac{A_3}{L} \Big)
 \end{aligned}
\end{equation}
and the corresponding line-element is
\begin{equation}
 \label{metricTsT3w3}
 \begin{aligned}
  ds^2 & = \frac{\zeta}{L^2} \left( r^2 \left( - dt^2 + \frac{dw^2}{W^2} + dz^2 + L^2 F \, d\phi^2 \right) + L^2 \frac{dr^2}{r^2 F \l^6} + L^4 d\th^2 \right)
  \\[5pt]
   & + \frac{L^2}{\zeta} \left( \cos^2\th \, d\psi^2 + \cos^2\th \, \sin^2\psi \Big( d\phi_1 + \frac{A_1}{L} \Big)^2 + \cos^2\th \, \cos^2\psi \Big( d\phi_2 + \frac{A_2}{L} \Big)^2 \right.
   \\[5pt]
   & \left. + \frac{\l^6 \sin^2\th}{W^2} \Big( d\phi_3 + \frac{A_3}{L} \Big)^2 \right) \, .
 \end{aligned}
\end{equation}
The NS sector is also supported by a non-trivial dilaton of the usual form \eqref{dilatonTsT121}, with $W$ given in eq. \eqref{frameDeformationTsT3w3}, and a two-form given by
\begin{equation}
 B_2 = \sqrt{W^2 - 1} \, e^1 \wedge e^9 \, .
\end{equation}
Setting $\g = 0$ results in a trivial $B_2$.

The RR sector contains only the three-form and the self-dual five-form. For the three-form, we obtain
\begin{equation}
 \label{F3TsT3w3}
 F_3 = \sqrt{W^2 - 1} \, \frac{\l^9 A'_{3 \phi}}{\zeta^{\frac{3}{2}}} \, \cos\th \, e^{025} \, .
\end{equation}
The five-form can be expressed as in eq. \eqref{F5ANO}, ensuring its self-duality, with
\begin{equation}
 \begin{aligned}
  G_5 & = \frac{2}{L} \frac{W \l^3 (1 + \zeta^2)}{\zeta^{\frac{5}{2}}} \, e^{01234}
            - \varepsilon \ell^2 \frac{W\sqrt{F}}{r^2 \zeta^{\frac{5}{2}}} \sin(2 \th) \, e^{01235}
            - \frac{W \, A'_{1\phi}}{\zeta^{\frac{3}{2}}} \sin\th \sin\psi \, e^{01257}
            \\[5pt]
           & - \frac{W A'_{1\phi}}{\zeta^{\frac{3}{2}}} \sin\th \cos\psi \, e^{01258}
              + \frac{W \, A'_{1\phi}}{\sqrt{\zeta}} \cos\psi \, e^{01267}
               - \frac{W A'_{1\phi}}{\sqrt{\zeta}} \sin\psi \, e^{01268}
            \\[5pt]
            & + \frac{\l^9 A'_{3\phi}}{\zeta^{\frac{3}{2}}} \cos\th \, e^{01259} \, .
 \end{aligned}
\end{equation}
In the absence of the deformation, only the five-form survives.


\section{Observables}

In this section we study various observables for each of the TsT-deformed backgrounds of Sec. \ref{SupergravityBackgrounds}. These include the Page charges, Wilson loops, 't Hooft loops, entanglement entropy, and the holographic central charge flow.


\subsection{Page Charges}

The first observable to consider will be the Page charges \cite{Marolf:2000cb} for the various backgrounds presented in Sec. \ref{SupergravityBackgrounds}. These are associated to the presence of $Dp$ branes and they are defined on a $(8-p)$-dimensional cycle $\S_{8-p}$, that is transverse to the brane, as
\footnote{
For convenience we work in the units where $\a' = g_s = 1$.
}
\begin{equation}
    Q_{Dp} = \frac{1}{(2 \pi)^{7 - p}} \int_{\S_{8-p}} \widehat{F}_{8-p} \, .
\end{equation}
For a type-IIB solution the various Page fluxes are defined as follows
\footnote{
For the RR forms of higher rank, we use their relation to the lower rank ones via the democratic formulation \cite{Bergshoeff:2001pv} $F_p = (-1)^{[\frac{p}{2}]} \star F_{10-p} \,\, (p \geq 5)$, where $[]$ stands for the integer part.
}

\begin{center}
 \begin{tabular}{c c}
 \textbf{Brane} & \textbf{Page flux} \\[5pt] 
 $D1$ & $\widehat{F}_7 = F_7 - F_5 \wedge B_2 + \frac{1}{2} F_3 \wedge B_2 \wedge B_2 - \frac{1}{6} F_1 \wedge B_2 \wedge B_2 \wedge B_2$ \\[5pt]  
 $D3$ & $\widehat{F}_5 = F_5 - F_3 \wedge B_2 + \frac{1}{2} F_1 \wedge B_2 \wedge B_2$ \\[5pt]
 $D5$ & $\widehat{F}_3 = F_3 - F_1 \wedge B_2$ \\[5pt]
 $D7$ & $\widehat{F}_1 = F_1$
\end{tabular}
\end{center}
Due to their quantised nature, the Page charges are expected to impose quantisation conditions on the various parameters that enter the supergravity solutions. Let us see this more explicitly.

\subsubsection*{The case of the seed background}

The RR sector of the undeformed solution of Sec. \ref{solutionANO} contains only the self-dual five-form, which is associated to the presence of $D3$ branes. The five-dimensional cycle transverse to the $D3$ branes is spanned by the directions $\S_{5} = (\th, \psi, \phi_1, \phi_2, \phi_3)$. To compute the Page charge, one has to look at the asymptotic behavior of the Page flux at large $r$, where the geometry approaches $AdS_5 \times S^5$. In this case
\begin{equation}
\label{D3PageFluxSeed}
    \widehat{F}_5\Big|_{\S_5} = - 2 L^4 \cos^3\th \sin\th \sin(2\psi) d\th \wedge d\psi \wedge d\phi_1 \wedge d\phi_2 \wedge d\phi_3 \, .
\end{equation}
The corresponding charge is
\begin{equation}
 \label{D3PageChargeANO}
    Q_{D3} = \frac{1}{(2\pi)^4} \int_{\S_5} \widehat{F}_5 = \frac{L^4}{4 \pi} = N \, \in \, \mathbb{N} \, .
\end{equation}

\subsubsection*{The case of the marginal deformation I}

The TsT transformation along $(\phi_1 , \phi_2)$ generates a RR three-form, suggesting the presence of $D5$ branes in addition to the $D3$ ones. It turns out that the $D3$ Page charge is not affected by the TsT transformation, and therefore is given by \eqref{D3PageChargeANO}. On the other hand, the $D5$ branes are transverse to the three-cycle spanned in the directions $\S_3 = (\th , \psi , \phi_3)$. In order to compute the corresponding Page charge, one has to look at the asymptotic behaviour of $\widehat{F}_3$ at large $r$
\begin{equation}
    \widehat{F}_3 \Big|_{\S_3} = - 2 \g L^4 \cos^3\th \sin\th \sin(2 \psi) d\th \wedge d\psi \wedge d\phi_3 \, .
\end{equation}
As a result
\begin{equation}
 \label{D5PageChargeMarginalI}
    Q_{D5} = \frac{1}{(2 \pi)^2} \int_{\S_3} \widehat{F}_3 = \g \frac{L^4}{4 \pi} = \g N = M \, \in \, \mathbb{N} \, .
\end{equation}
The last provides a good quantisation condition for the $D5$ branes, provided that the deformation parameter $\g$ is a rational number. Such quantisation conditions have appeared in similar setups; see, for example, \cite{Castellani:2024pmx}.

\subsubsection*{The case of the marginal deformation II}

As in the previous case, in addition to the self-dual five-form, there is also a three-form flux generated by the TsT transformation along the directions $(\phi_1, \phi_3)$. This flux is associated with the presence of a stack of $D5$ branes, which are transverse to the three-cycle spanned by the directions $\S'_3 = (\th, \psi, \phi_2)$. Again, the Page charge for the stack of $D3$ branes, induced by the RR five-form, is not affected by the TsT transformation. Therefore, it satisfies the quantisation condition \eqref{D3PageChargeANO}. However, looking at the asymptotic behaviour of $\widehat{F}_3$ at large $r$ we obtain
\begin{equation}
    \widehat{F}_3 \Big|_{\S'_3} = - 2 \g L^4 \cos^3\th \sin\th \sin(2 \psi) d\th \wedge d\psi \wedge d\phi_2 \, .
\end{equation}
This implies the quantisation condition
\begin{equation}
    Q'_{D5} = \frac{1}{(2 \pi)^2} \int_{\S'_3} \widehat{F}_3 = \g \frac{L^4}{4 \pi} = \g N = M \, \in \, \mathbb{N} \, ,
\end{equation}
which is the same as in \eqref{D5PageChargeMarginalI}, and makes sense when $\g$ is a rational number.

\subsubsection*{The case of the dipole deformations I \& II}

The backgrounds of Secs. \ref{solutionTsT1w1} and \ref{solutionTsT3w3} support only the Page charge associated with $D3$ branes. This is defined on the five-cycle $\S_{5} = (\th, \psi, \phi_1, \phi_2, \phi_3)$, where the large-$r$ behaviour of the associated Page flux is as in eq. \eqref{D3PageFluxSeed}, and the corresponding charge is given in eq. \eqref{D3PageChargeANO}.

\subsection{Wilson Loop}

We now turn to the computation of the energy between a quark--anti-quark pair as a function of their separation. This is realised holographically by a string that extends along the time-like direction and one of the external spatial directions of the ten-dimensional geometry, with a non-trivial profile in the radial coordinate. We illustrate this explicitly for each of the geometries derived after a TsT transformation and presented in Sec.~\ref{SupergravityBackgrounds}.

\subsubsection*{The case of the marginal deformations I \& II}
\label{WLmarginalDef}

We consider a string embedded in the geometries \eqref{metricTsT121} and \eqref{metricTsT131}, parametrised as
\begin{equation}
\label{WLansatz}
    \tau = t \, , \qquad \sigma = w \, , \qquad r = r(\sigma) \, ,
\end{equation}
with all remaining coordinates kept fixed. Such a configuration is consistent when $\th = 0$ or $\th = \frac{\pi}{2}$. Moreover, the string extends from $\sigma = - \frac{d}{2}$ to $\sigma = \frac{d}{2}$ at the boundary, where
\begin{equation}
\label{StringBCsMarginal}
    r\big( \pm \nicefrac{d}{2} \big) = \infty \, ,
\end{equation}
and it reaches the turning point $r = r_0$ in the bulk. We expect the corresponding Wilson loop to exhibit the same behaviour as in the case of the seed geometry \eqref{metricANO}, which was studied in \cite{Chatzis:2025dnu,Chatzis:2025hek}. This is because the string probes only the external part of the ten-dimensional geometry, which is identical in all three cases, namely \eqref{metricANO}, \eqref{metricTsT121}, and \eqref{metricTsT131}.

The induced metric of the string reads
\begin{equation}
    ds^2_{\text{ind}} = - \frac{r^2 \zeta(r, \th_0)}{L^2} d\tau^2 + \Bigg( \frac{r^2 \zeta(r, \th_0)}{L^2} + \frac{r'^2 \zeta(r, \th_0)}{r^2 F(r) \lambda(r)^6} \Bigg) d\sigma^2 \, , \qquad
    \th_0 = 0 \, , \frac{\pi}{2} \, ,
\end{equation}
where the prime stands for derivative with respect to $\sigma$. The corresponding Nambu-Goto action is
\begin{equation}
\label{NGactionWL}
    S_\text{NG} = \frac{\cT}{2 \pi} \int\limits_{- \nicefrac{d}{2}}^{\nicefrac{d}{2}} d\sigma \, \sqrt{\cF^2 + \cG^2 \, r'^2} \, ,
\end{equation}
where
\begin{equation}
    \cF = \frac{r^2 \zeta(r, \th_0)}{L^2} \, , \qquad
    \cG = \frac{\zeta(r, \th_0)}{L \sqrt{F(r)} \lambda(r)^3} \, ,
\end{equation}
and $\cT = \int d\tau$. The distance $d$, which corresponds to the separation of the quark-anti-quark pair, can be expressed as a function of the turning point through the formula
\begin{equation}
\label{qqbSeparation}
    d(r_0) = 2 \cF_0 \int\limits_{r_0}^{\infty} dr \, \frac{\cG}{\cF} \frac{1}{\sqrt{\cF^2 - \cF^2_0}} \, ,
\end{equation}
where $\cF_0 := \left. \cF \right|_{r = r_0}$. A similar expression holds for the energy of the quark-anti-quark pair, where now
\begin{equation}
\label{qqbEnergy}
    \cE(r_0) = \frac{1}{\pi} \int\limits_{r_0}^{\infty} dr \, \frac{\cF \cG}{\sqrt{\cF^2 - \cF^2_0}} - \frac{1}{\pi} \int\limits_{r_\star}^{\infty} dr \, \cG \, .
\end{equation}
The last term in the expression above corresponds to the energy of two strings that stretch between the boundary and the end of the space $r = r_\star$.

\begin{figure}[ht]
    \centering
    \begin{subfigure}[b]{0.495\textwidth}
        \centering
        \includegraphics[width=\linewidth]{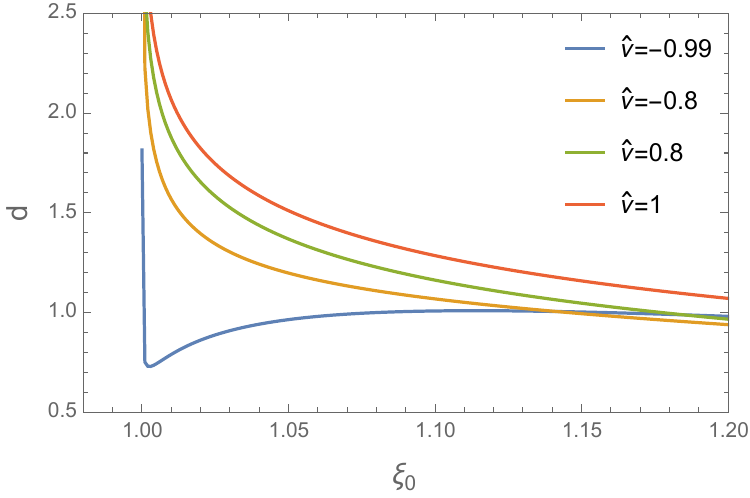}
        \caption{Separation vs turning point for $\th_0 = 0$.}
        \label{fig:WLimage1}
    \end{subfigure}
    \hfill
    \begin{subfigure}[b]{0.495\textwidth}
        \centering
        \includegraphics[width=\linewidth]{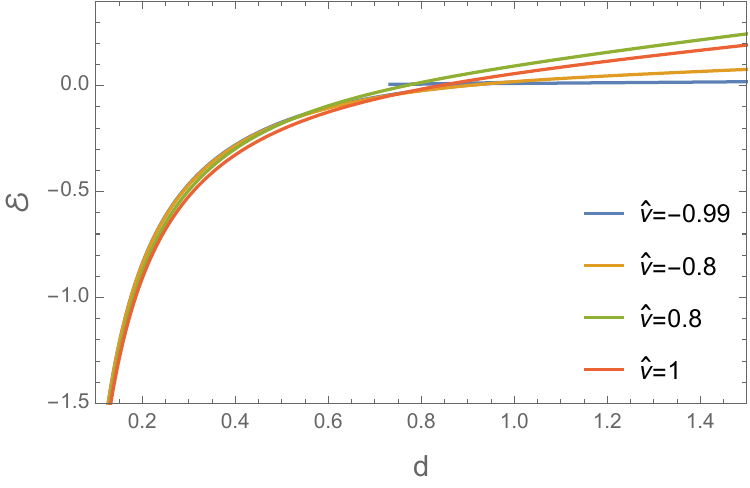}
        \caption{Energy vs separation for $\th_0 = 0$.}
        \label{fig:WLimage2}
    \end{subfigure}
    
    \vspace{0.5cm} 
    
    \begin{subfigure}[b]{0.495\textwidth}
        \centering
        \includegraphics[width=\linewidth]{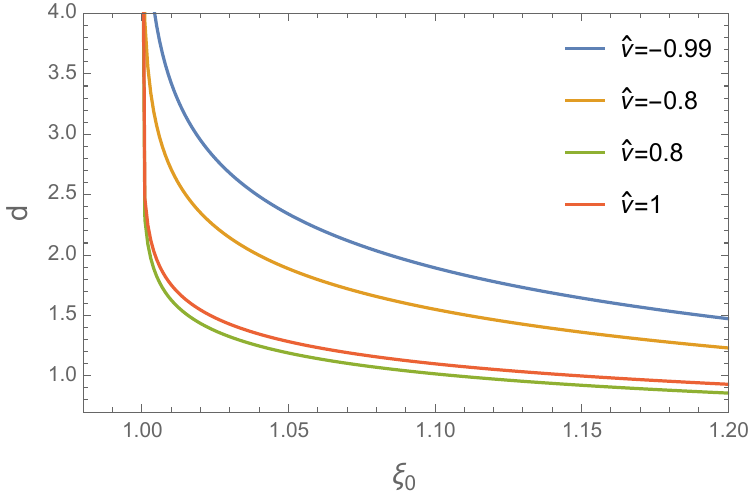}
        \caption{Separation vs turning point for $\th_0 = \frac{\pi}{2}$.}
        \label{fig:WLimage3}
    \end{subfigure}
    \hfill
    \begin{subfigure}[b]{0.495\textwidth}
        \centering
        \includegraphics[width=\linewidth]{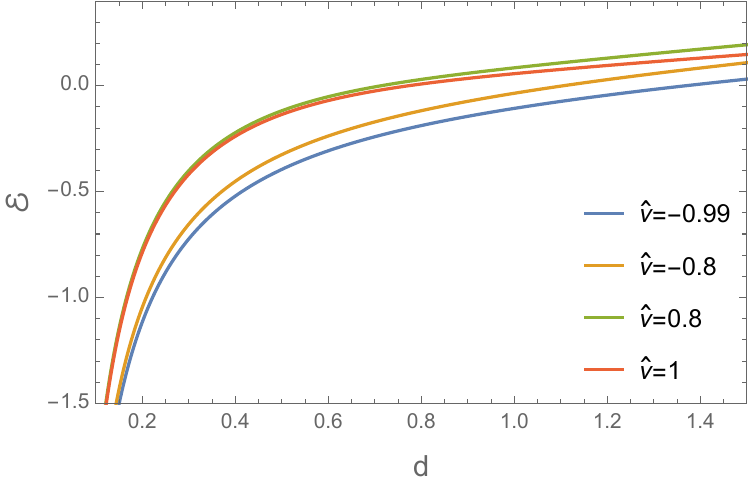}
        \caption{Energy vs separation for $\th_0 = \frac{\pi}{2}$.}
        \label{fig:WLimage4}
    \end{subfigure}
    
    \caption{The separation length as a function of the turning point, and the quark-anti-quark potential as a function of the separation, for different values of $\hat{\nu}$. For simplicity we set $L = \ell = 1$. The figures at the top correspond to the embedding with $\th_0 = 0$, while the ones at the bottom to the embedding with $\th_0 = \frac{\pi}{2}$.}
    \label{fig:WLmarginal}
\end{figure}

\begin{figure}[h ]
    \centering
    \includegraphics[width=0.7\textwidth]{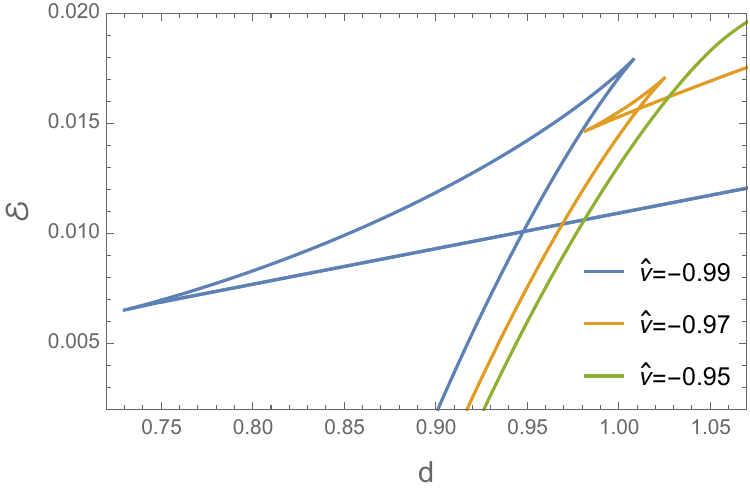}
    \caption{The energy as a function of the separation when $\th_0 = 0$ and $\hat{\nu}$ takes values near $-1$. The triangle disappears near $\hat{\nu} \simeq - 0.95$. For simplicity we set $L = \ell = 1$.}
    \label{fig:triangleWLmarginal}
\end{figure}

The behaviour of the separation length of the quark--anti-quark pair as a function of the turning point $\xi_0$\footnote{It is more convenient to work with the coordinate $\xi = \frac{r}{r_\star}$.}, for each embedding, is illustrated in Figs.~\ref{fig:WLimage1} and \ref{fig:WLimage3}. One can observe that, in the case $\theta_0 = 0$, the separation length is not a one-to-one function of $\xi_0$ for values of $\hat{\nu}$ close to $-1$. This feature is also reflected in the quark--anti-quark potential shown in Figs.~\ref{fig:WLimage2} and \ref{fig:triangleWLmarginal}, pointing towards a first-order phase transition. We believe that this behaviour is a manifestation of the fact that the supergravity approximation cannot be trusted near $\theta_0 = 0$ when the parameter $\hat{\nu}$ approaches the value $-1$. Otherwise, the quark--anti-quark potential exhibits Coulombic behaviour at small separations (due to conformality in the UV) and grows linearly at large separations, signalling confinement in the IR.

\subsubsection*{The case of the dipole deformation I}
\label{WLdipoleDefI}

We now move to the study of the Wilson loop, corresponding to a string embedded in the geometry of eq. \eqref{metricTsT1w1} according to the ansatz in eq. \eqref{WLansatz}. In this case, the embedding is consistent provided that $\th = \frac{\pi}{2}$, or $\th = 0$ with $\psi = 0$ or $\psi = \frac{\pi}{2}$. The induced metric now becomes
\begin{equation}
    ds^2_{\text{ind}} = - \frac{r^2 \zeta(r, \th_0)}{L^2} d\tau^2 + \Bigg( \frac{r^2 \zeta(r, \th_0)}{L^2 W(r, \th_0 , \psi_0)^2} + \frac{r'^2 \zeta(r, \th_0)}{r^2 F(r) \lambda(r)^6} \Bigg) d\sigma^2 \, ,
\end{equation}
with the function $W$ given in eq. \eqref{frameDeformationTsT1w1}. The constants $\th_0$ and $\psi_0$ take values according to the following three cases
\begin{equation}
\label{WLembeddingsDipoleI}
    \textbf{(A)} \quad \th_0 = \frac{\pi}{2} \, , \psi_0 = \text{any const.}, \quad
    \textbf{(B)} \quad \th_0 = \psi_0 = 0 \, , \quad
    \textbf{(C)} \quad \th_0 = 0 \, , \psi_0 = \frac{\pi}{2} \, .
\end{equation}
The Nambu--Goto action again takes the usual form of eq. \eqref{NGactionWL}, where now $\cF$ is dressed with the function $W$, i.e.
\begin{equation}
    \cF = \frac{r^2 \zeta(r, \th_0)}{L^2 W(r, \th_0, \psi_0)} \, , \qquad
    \cG = \frac{\zeta(r, \th_0)}{L \sqrt{F(r)} \lambda(r)^3} \, .
\end{equation}
Notice that in cases \textbf{(A)} and \textbf{(B)} of eq. \eqref{WLembeddingsDipoleI}, the function $W$ takes the value one. Therefore, we expect the Wilson loop in these cases to behave as in the original or marginally deformed backgrounds studied in \cite{Chatzis:2025dnu,Chatzis:2025hek} and Sec.~\ref{WLmarginalDef}. However, case \textbf{(C)} turns out to be different, and the Wilson loop is affected by the deformation parameter $\g$ as we will see shortly.

\begin{figure}[h]
    \centering
    \begin{subfigure}[b]{0.495\textwidth}
        \centering
        \includegraphics[width=\linewidth]{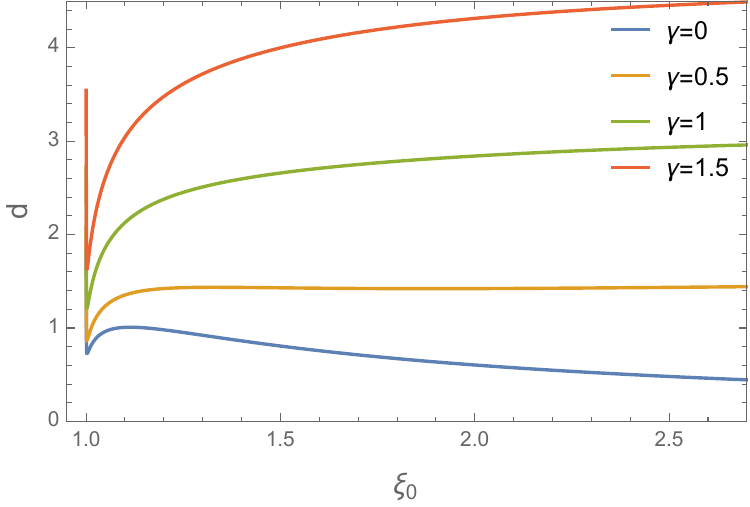}
        \caption{Separation vs turning point for $\hat{\nu} = -0.99$.}
        \label{fig:dipoleIlengthWLimage1}
    \end{subfigure}
    \hfill
    \begin{subfigure}[b]{0.495\textwidth}
        \centering
        \includegraphics[width=\linewidth]{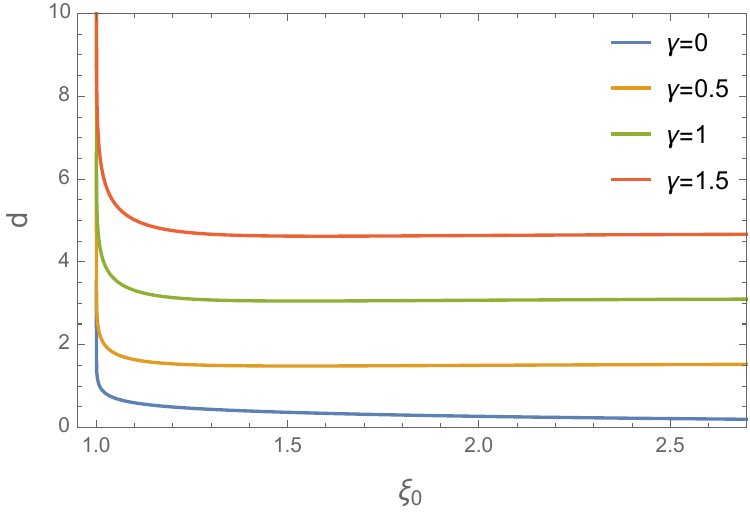}
        \caption{Separation vs turning point for $\hat{\nu} = -0.2$.}
        \label{fig:dipoleIlengthWLimage2}
    \end{subfigure}
    
    \vspace{0.5cm} 
    
    \begin{subfigure}[b]{0.495\textwidth}
        \centering
        \includegraphics[width=\linewidth]{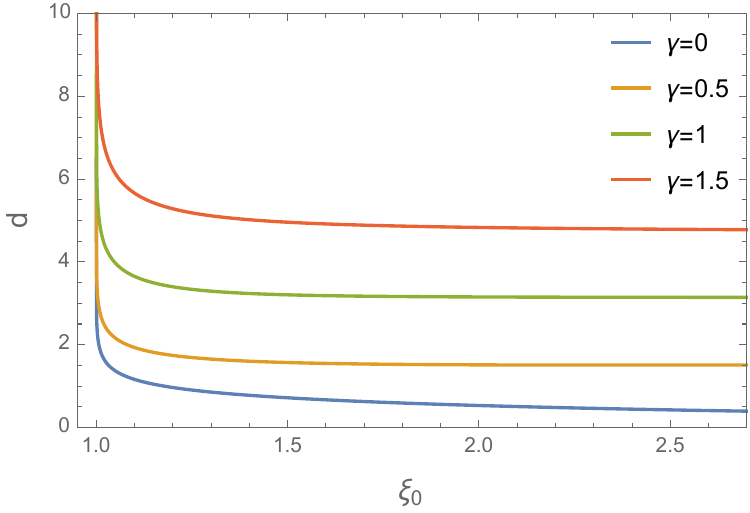}
        \caption{Separation vs turning point for $\hat{\nu} = 0.8$.}
        \label{fig:dipoleIlengthWLimage3}
    \end{subfigure}
    \hfill
    \begin{subfigure}[b]{0.495\textwidth}
        \centering
        \includegraphics[width=\linewidth]{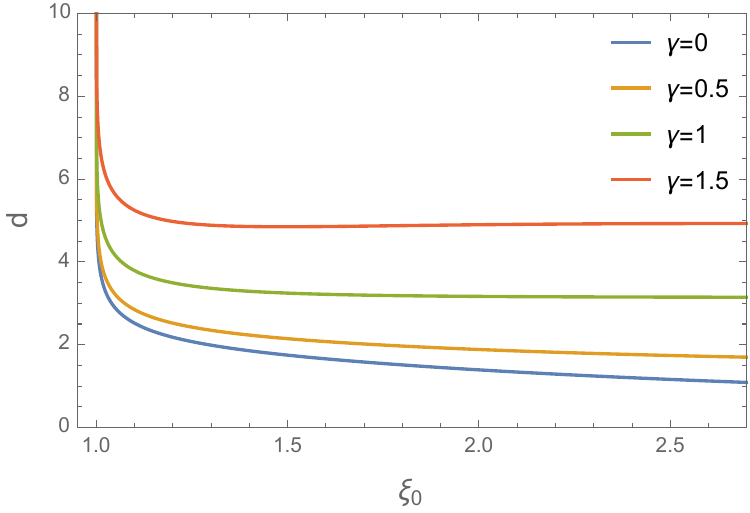}
        \caption{Separation vs turning point for $\hat{\nu} = 7$.}
        \label{fig:dipoleIlengthWLimage4}
    \end{subfigure}
    
    \caption{The quark-anti-quark separation as a function of the turning point when $\th_0 = 0$ and $\psi_0 = \nicefrac{\pi}{2}$. Each figure corresponds to a different value of $\hat{\nu}$ and each curve to a different value of $\g$. For simplicity we set $L = \ell = 1$.}
    \label{fig:WLdipoleIlengthFixedn}
\end{figure}

\begin{figure}[h]
    \centering
    \begin{subfigure}[b]{0.495\textwidth}
        \centering
        \includegraphics[width=\linewidth]{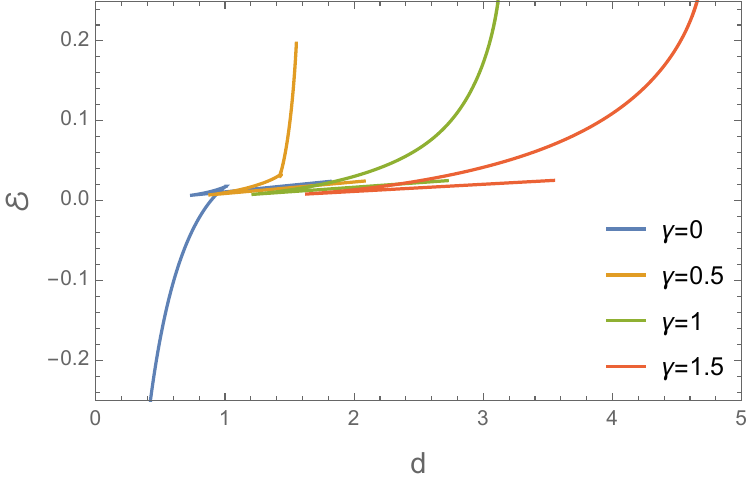}
        \caption{Energy vs separation for $\hat{\nu} = -0.99$.}
        \label{fig:dipoleIenergyWLimage1}
    \end{subfigure}
    \hfill
    \begin{subfigure}[b]{0.495\textwidth}
        \centering
        \includegraphics[width=\linewidth]{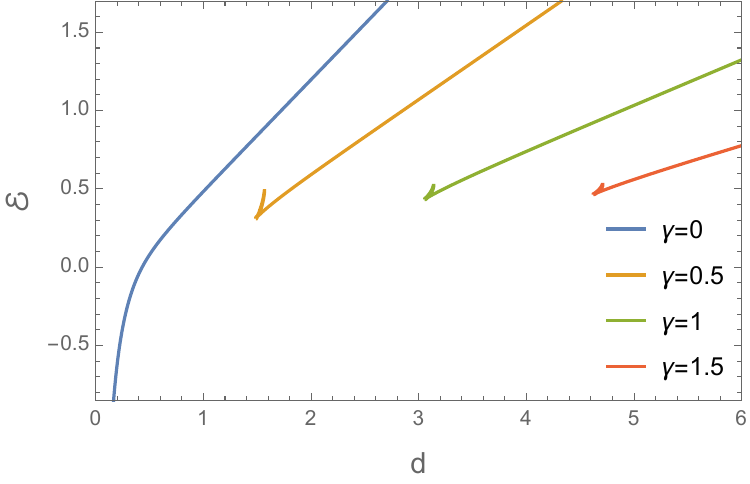}
        \caption{Energy vs separation for $\hat{\nu} = -0.2$.}
        \label{fig:dipoleIenergyWLimage2}
    \end{subfigure}
    
    \vspace{0.5cm} 
    
    \begin{subfigure}[b]{0.495\textwidth}
        \centering
        \includegraphics[width=\linewidth]{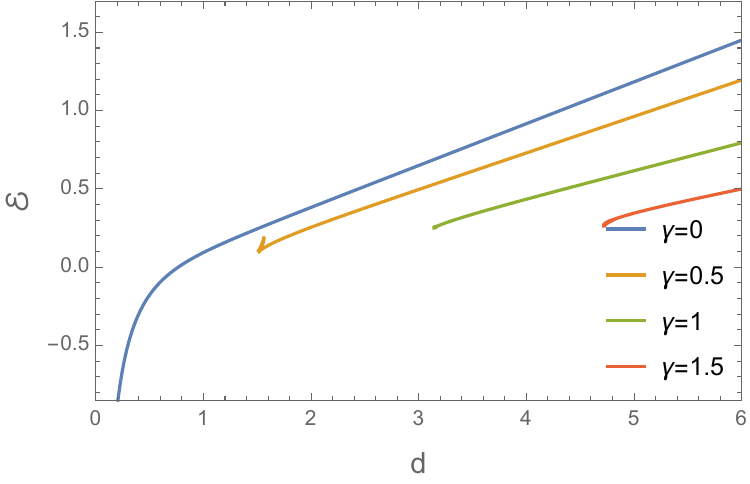}
        \caption{Energy vs separation for $\hat{\nu} = 0.8$.}
        \label{fig:dipoleIenergyWLimage3}
    \end{subfigure}
    \hfill
    \begin{subfigure}[b]{0.495\textwidth}
        \centering
        \includegraphics[width=\linewidth]{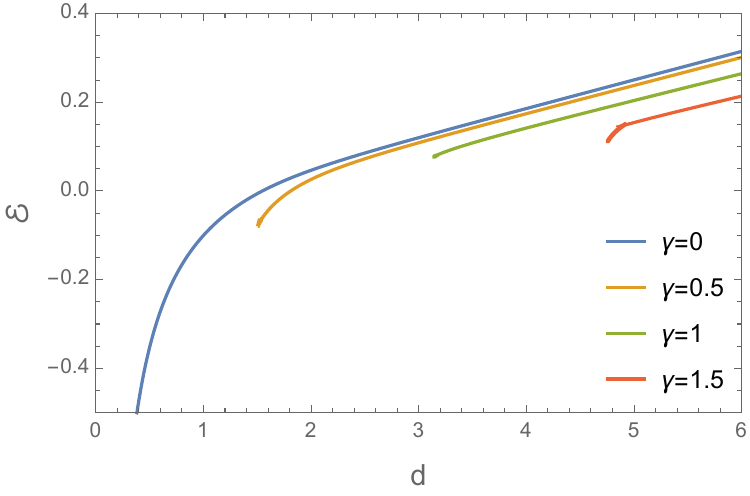}
        \caption{Energy vs separation for $\hat{\nu} = 7$.}
        \label{fig:dipoleIenergyWLimage4}
    \end{subfigure}
    
    \caption{The energy as a function of the separation when $\th_0 = 0$ and $\psi_0 = \nicefrac{\pi}{2}$. Each figure corresponds to a different value of $\hat{\nu}$ and each curve to a different value of $\g$. For simplicity we set $L = \ell = 1$.}
    \label{fig:WLdipoleIenergyFixedn}
\end{figure}

The separation of the quark-anti-quark pair and the corresponding energy are again given by the formulas \eqref{qqbSeparation} and \eqref{qqbEnergy}, respectively. Their behaviour in the case of interest, i.e.\ case \textbf{(C)}, is illustrated in Figs. \ref{fig:WLdipoleIlengthFixedn} and \ref{fig:WLdipoleIenergyFixedn}. In the absence of deformation, i.e.\ when $\g = 0$, the function $W$ takes the value one, and we therefore expect to recover the same behaviour as in the seed and marginally deformed backgrounds. Figure \ref{fig:WLdipoleIlengthFixedn} reveals that the quark-anti-quark separation is a double-valued function with a \emph{sharp} local minimum for negative values of $\hat{\nu}$, approaching $\hat{\nu} = -1$. We believe that this is an artefact due to the singularity at $\xi = 1, \, \th = 0$ when $\hat{\nu} = -1$, which is also reflected in the energy plots through the appearance of a triangle indicating a phase transition. As a consequence, this behaviour should not be trusted. 

\begin{figure}[h]
    \centering
    \begin{subfigure}[b]{0.495\textwidth}
        \centering
        \includegraphics[width=\linewidth]{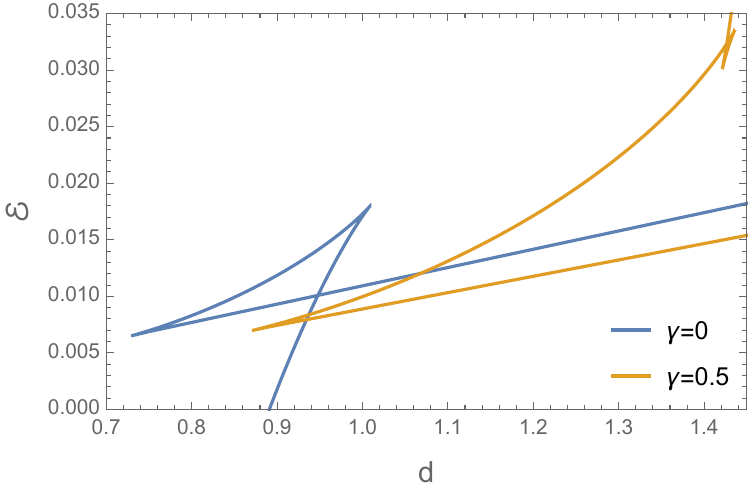}
        \captionsetup{width=.6\linewidth,justification=centering}
        \caption{Energy vs separation for $\hat{\nu} = -0.99$ and $\g = 0 \, \& \, 0.5$.}
        \label{fig:dipoleIenergyWLimage9}
    \end{subfigure}
    \hfill
    \begin{subfigure}[b]{0.495\textwidth}
        \centering
        \includegraphics[width=\linewidth]{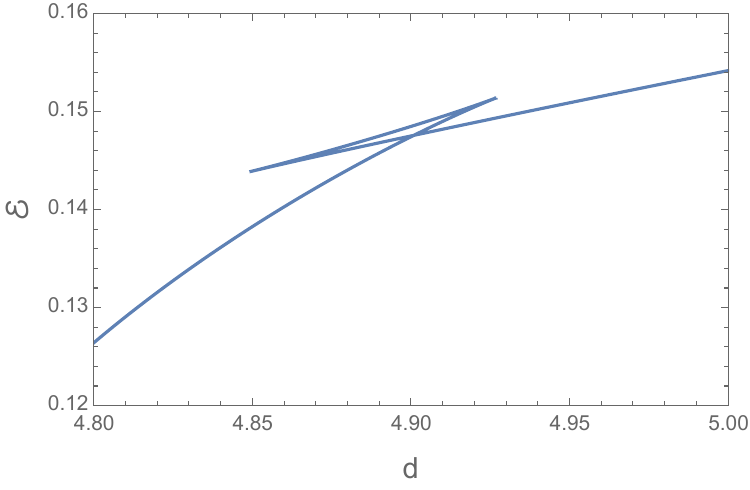}
        \captionsetup{width=.6\linewidth,justification=centering}
        \caption{Energy vs separation for $\hat{\nu} = 7$ and $\g = 1.5$.}
        \label{fig:dipoleIenergyWLimage10}
    \end{subfigure}
    
    \caption{Zoom-ins of Fig. \ref{fig:dipoleIenergyWLimage1} \textbf{(left)} and Fig. \ref{fig:dipoleIenergyWLimage4} \textbf{(right)} reveal the triangular shape.}
    \label{fig:WLdipoleIenergyTriangles}
\end{figure}

Figure \ref{fig:WLdipoleIenergyFixedn} exhibits a \emph{wedge} in the energy for small separations when the deformation is non-zero. This wedge coincides with the absence of conformality in the UV and tends to disappear as $\hat{\nu}$ or $\g$ are increased. For sufficiently large separations, the energy increases linearly, pointing towards a confining behaviour in the IR. A closer inspection of Fig. \ref{fig:WLdipoleIenergyFixedn} shows that, for $\hat{\nu} = -0.99$, the appearance of the triangular structure in the energy competes with increasing deformation, whereas the opposite behaviour is observed for larger values of $\hat{\nu}$, such as $\hat{\nu} = 7$. The latter can be interpreted as a phase transition. These triangular structures are illustrated more clearly in Fig. \ref{fig:WLdipoleIenergyTriangles}.

\subsubsection*{The case of the dipole deformation II}

In the last example, we again study the Wilson loop corresponding to the string described by the ansatz of eq.~\eqref{WLansatz}, now embedded in the geometry of eq.~\eqref{metricTsT3w3}. This embedding is consistent provided that $\th$ is fixed to the values $0$ or $\nicefrac{\pi}{2}$. The induced metric on the string then reads
\begin{equation}
    ds^2_{\text{ind}} = - \frac{r^2 \zeta(r, \th_0)}{L^2} d\tau^2 
    + \Bigg( \frac{r^2 \zeta(r, \th_0)}{L^2 W(r, \th_0)^2} 
    + \frac{r'^2 \zeta(r, \th_0)}{r^2 F(r) \lambda(r)^6} \Bigg) d\sigma^2 \, ,
\end{equation}
where the function $W$ is given in eq.~\eqref{frameDeformationTsT3w3} and $\th_0 = 0, \, \nicefrac{\pi}{2}$. The Nambu--Goto action takes the usual form of eq.~\eqref{NGactionWL}, with the functions $\cF$ and $\cG$ given by
\begin{equation}
    \cF = \frac{r^2 \zeta(r, \th_0)}{L^2 W(r, \th_0)} \, , \qquad
    \cG = \frac{\zeta(r, \th_0)}{L \sqrt{F(r)} \lambda(r)^3} \, .
\end{equation}
Notice that when $\th_0 = 0$ the function $W$ takes the value one. Therefore, we expect the Wilson loop to behave as in the original and marginally deformed backgrounds discussed in \cite{Chatzis:2025dnu,Chatzis:2025hek} and Sec.~\ref{WLmarginalDef}. For this reason, we focus on the case $\th_0 = \nicefrac{\pi}{2}$, which captures the dependence on the deformation parameter.

\begin{figure}[h]
    \centering
    \begin{subfigure}[b]{0.495\textwidth}
        \centering
        \includegraphics[width=\linewidth]{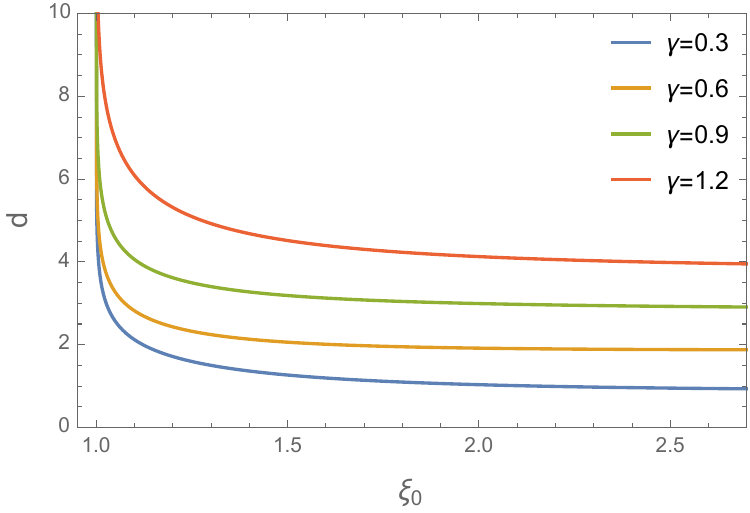}
        \caption{Separation vs turning point for $\hat{\nu} = -0.99$.}
        \label{fig:dipoleIIlengthWLimage1}
    \end{subfigure}
    \hfill
    \begin{subfigure}[b]{0.495\textwidth}
        \centering
        \includegraphics[width=\linewidth]{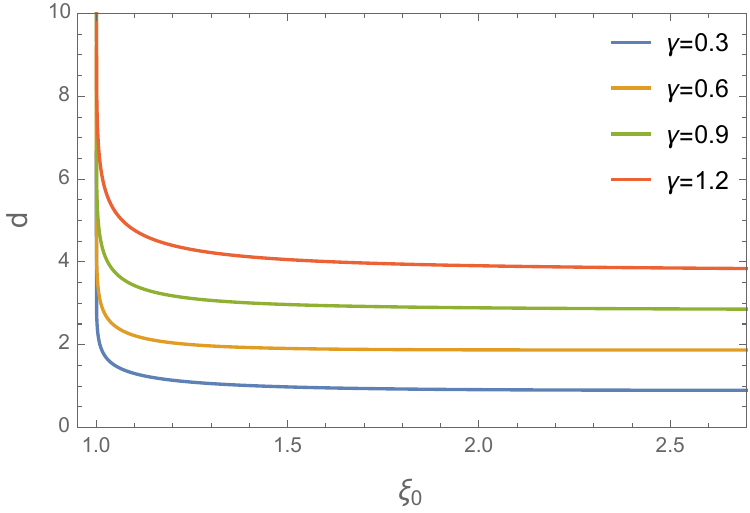}
        \caption{Separation vs turning point for $\hat{\nu} = -0.4$.}
        \label{fig:dipoleIIlengthWLimage2}
    \end{subfigure}
    
    \vspace{0.5cm} 
    
    \begin{subfigure}[b]{0.495\textwidth}
        \centering
        \includegraphics[width=\linewidth]{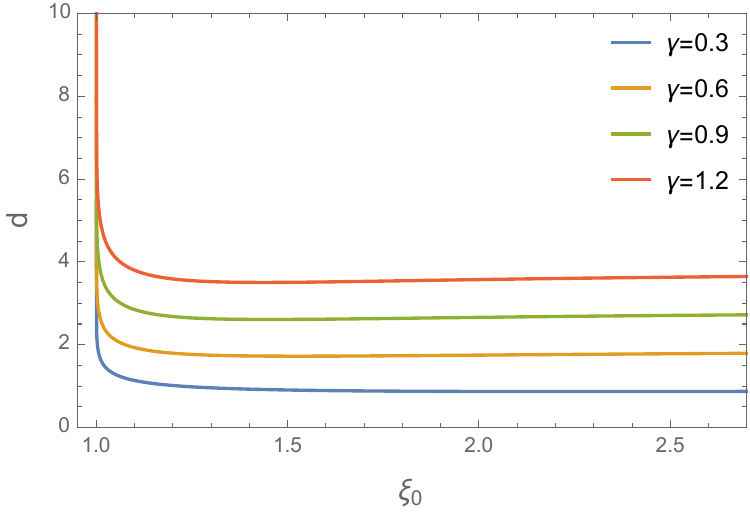}
        \caption{Separation vs turning point for $\hat{\nu} = 0.4$.}
        \label{fig:dipoleIIlengthWLimage3}
    \end{subfigure}
    \hfill
    \begin{subfigure}[b]{0.495\textwidth}
        \centering
        \includegraphics[width=\linewidth]{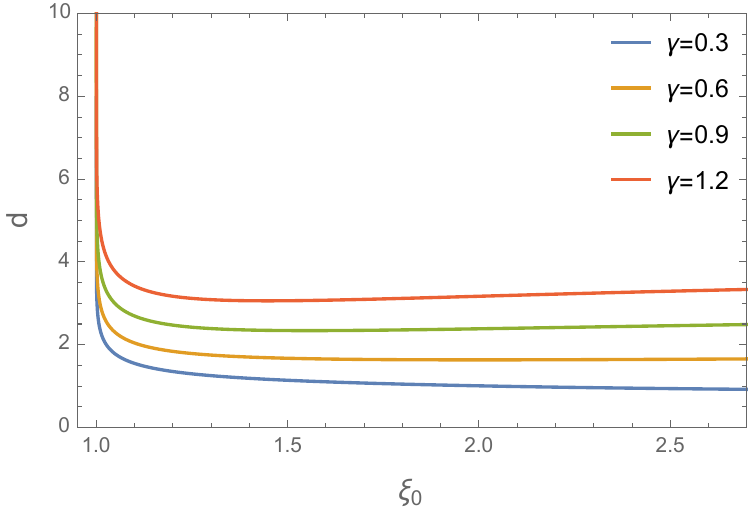}
        \caption{Separation vs turning point for $\hat{\nu} = 2$.}
        \label{fig:dipoleIIlengthWLimage4}
    \end{subfigure}
    
    \caption{The quark-anti-quark separation as a function of the turning point when $\th_0 = \nicefrac{\pi}{2}$. Each figure corresponds to a different value of $\hat{\nu}$ and each curve to a different value of $\g$. For simplicity we set $L = \ell = 1$.}
    \label{fig:WLdipoleIIlengthFixedn}
\end{figure}

\begin{figure}[h]
    \centering
    \begin{subfigure}[b]{0.495\textwidth}
        \centering
        \includegraphics[width=\linewidth]{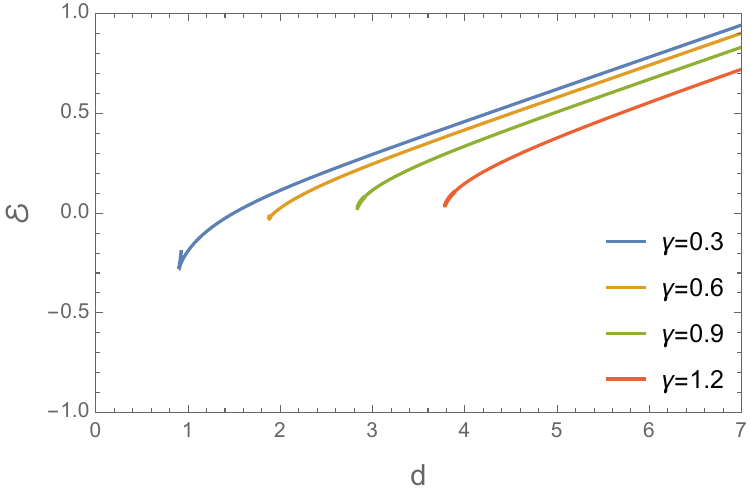}
        \caption{Energy vs separation for $\hat{\nu} = -0.99$.}
        \label{fig:dipoleIIenergyWLimage1}
    \end{subfigure}
    \hfill
    \begin{subfigure}[b]{0.495\textwidth}
        \centering
        \includegraphics[width=\linewidth]{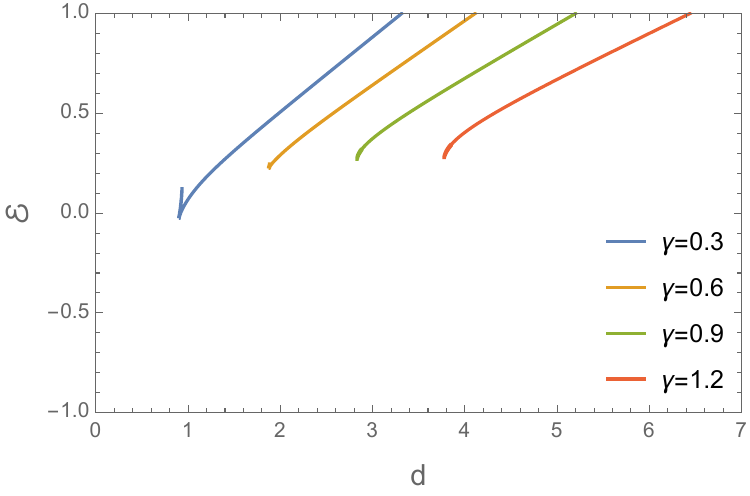}
        \caption{Energy vs separation for $\hat{\nu} = -0.4$.}
        \label{fig:dipoleIIenergyWLimage2}
    \end{subfigure}
    
    \vspace{0.5cm} 
    
    \begin{subfigure}[b]{0.495\textwidth}
        \centering
        \includegraphics[width=\linewidth]{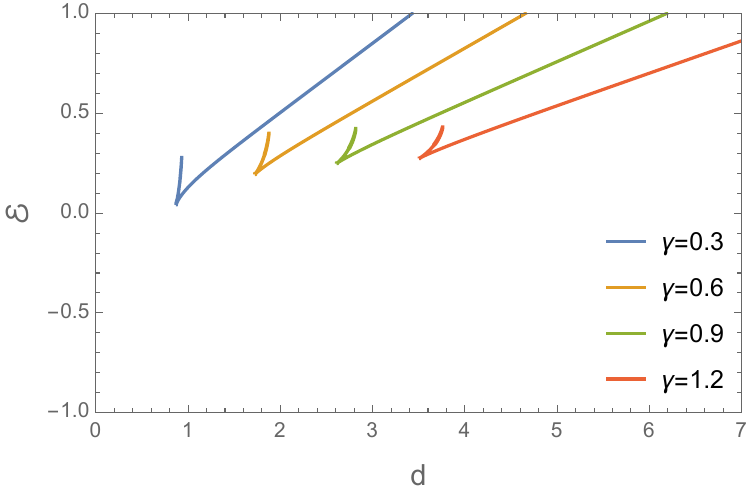}
        \caption{Energy vs separation for $\hat{\nu} = 0.4$.}
        \label{fig:dipoleIIenergyWLimage3}
    \end{subfigure}
    \hfill
    \begin{subfigure}[b]{0.495\textwidth}
        \centering
        \includegraphics[width=\linewidth]{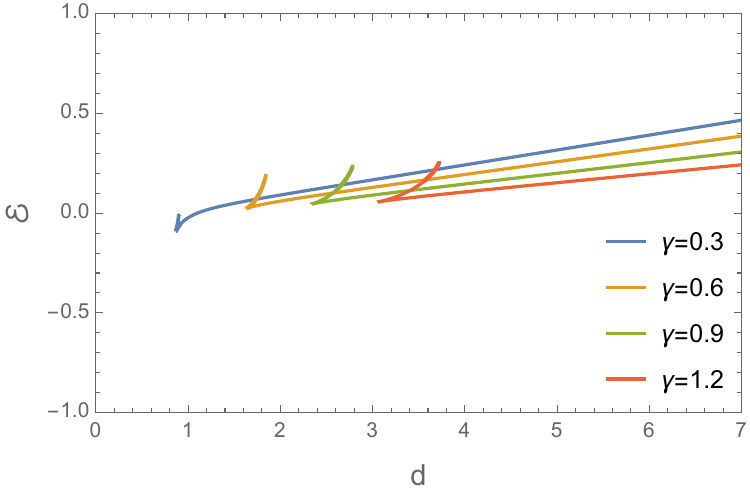}
        \caption{Energy vs separation for $\hat{\nu} = 2$.}
        \label{fig:dipoleenergyIIWLimage4}
    \end{subfigure}
    
    \caption{The energy as a function of the separation when $\th_0 = \nicefrac{\pi}{2}$. Each figure corresponds to a different value of $\hat{\nu}$ and each curve to a different value of $\g$. For simplicity we set $L = \ell = 1$.}
    \label{fig:WLdipoleIIenergyFixedn}
\end{figure}

The quark-anti-quark separation and the corresponding energy are again determined by the formulas \eqref{qqbSeparation} and \eqref{qqbEnergy}, respectively. Their behaviour is shown in Figs.~\ref{fig:WLdipoleIIlengthFixedn} and \ref{fig:WLdipoleIIenergyFixedn}. The dependence of the separation on the turning point $\xi_0$ is qualitatively similar to that of the example discussed in Sec.~\ref{WLdipoleDefI}. For the energy, we again observe the emergence of a wedge shape for small separations, where conformality is broken by the presence of the deformation, as well as a linear growth towards the IR, indicating confinement.

\subsection{'t Hooft Loop}

The next observable we would like to study is the 't Hooft loop, describing a magnetic monopole-anti-monopole pair. Holographically this is realised in terms of a $D3$ brane that extends in three external coordinates and wraps a circle in the internal space of the ten-dimensional geometry. Let us see this more explicitly for each of the backgrounds in Sec. \ref{SupergravityBackgrounds} found via a TsT transformation.

\subsubsection*{The case of the marginal deformations I \& II and the dipole deformation I}

The setup we consider here is that of a $D3$ brane which extends in the directions $(t, \, w, \, \phi, \, \phi_3)$ of the geometries \eqref{metricTsT121}, \eqref{metricTsT131} and \eqref{metricTsT1w1}. We also assume that $w$ takes values in $[- \nicefrac{d}{2} , \nicefrac{d}{2}]$ and a profile in the $r$-direction such that $r = r(w)$ and boundary conditions \eqref{StringBCsMarginal}. The rest of the coordinates are kept fixed and the configuration is consistent as long as the value of $\th$ is $\frac{\pi}{2}$, corresponding to $\zeta = W = 1$ for all three solutions. The dynamics of this $D3$ brane was discussed for the seed solution \ref{solutionANO} in \cite{Chatzis:2025hek}.

The induced metric on the above $D3$ brane reads
\begin{equation}
    \begin{aligned}
  ds^2 & = - \frac{r^2}{L^2} dt^2
  + \left( \frac{r^2}{L^2}
  + \frac{r'^2}{r^2 F \l^6} \right) dw^2
  + r^2 F \, d\phi^2
  + L^2 \l^6 \Big( d\phi_3 + \frac{A_3}{L} \Big)^2 \, .
 \end{aligned}
\end{equation}
Since $W = 1$ in all three cases, the dilaton and NS two-form do not contribute to the Dirac-Born-Infeld (DBI) action of the brane. Similarly the Wess-Zumino action is trivial and therefore the dynamics of the brane is described by
\begin{equation}
\label{tHooftD3action}
    S_{D3} = 2 \pi \, T_{D3} L_{\phi} \cT \int\limits_{- \nicefrac{d}{2}}^{\nicefrac{d}{2}} dw \sqrt{\cF^2 + \cG^2 \, r'^2} \, ,
\end{equation}
with
\begin{equation}
\label{FGtHooftMarginal}
    \cF = \frac{r^3 \l^3 \sqrt{F}}{L} \, , \qquad \cG = r \, .
\end{equation}
In the above expression $\cT = \int dt$ and the periodicity of $\phi$ is given in \eqref{phiPeriodicity}.

\begin{figure}[ht]
    \centering
    \begin{subfigure}[b]{0.495\textwidth}
        \centering
        \includegraphics[width=\linewidth]{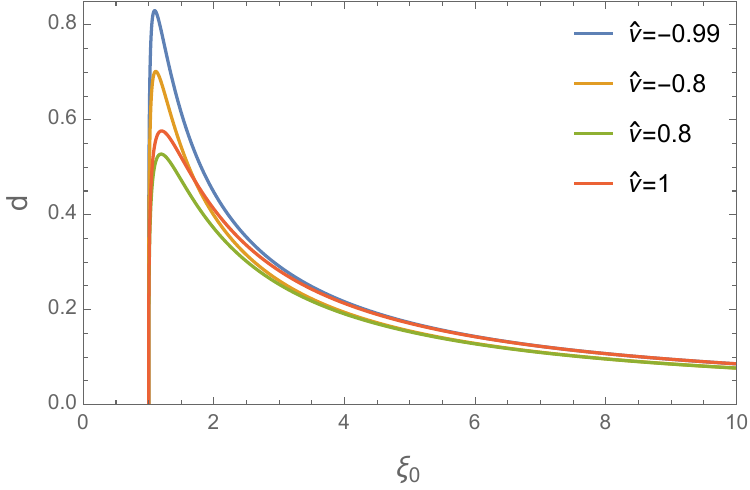}
        \caption{Separation vs turning point for $\th_0 = \frac{\pi}{2}$.}
        \label{fig:tHooftimage1}
    \end{subfigure}
    \hfill
    \begin{subfigure}[b]{0.495\textwidth}
        \centering
        \includegraphics[width=\linewidth]{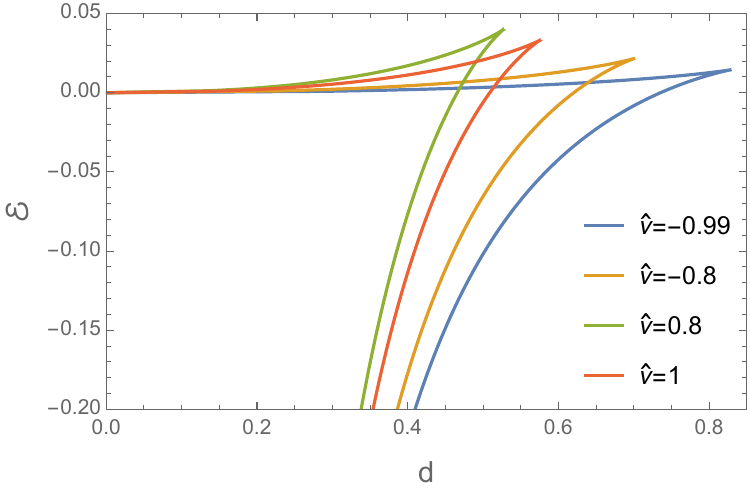}
        \caption{Energy vs separation for $\th_0 = \frac{\pi}{2}$.}
        \label{fig:tHooftimage2}
    \end{subfigure}
    
    \caption{The separation length as a function of the turning point, and the energy of the monopole-anti-monopole pair as a function of the separation, for different values of $\hat{\nu}$. For simplicity we set $L = \ell = 1$.}
    \label{fig:tHooftMarginal}
\end{figure}

The separation and energy of the magnetic monopole-anti-monopole pair are expressed as functions of the turning point $\xi_0$ through the formulas \eqref{qqbSeparation} and \eqref{qqbEnergy}, respectively. Since, for this embedding, the function $W$ takes the value $1$, we find that the separation and the energy are not affected by the deformation parameter $\gamma$. Therefore, their behaviour is expected to be the same as in the case of the undeformed background \eqref{solutionANO}, which was studied in \cite{Chatzis:2025hek}. In Fig.~\ref{fig:tHooftMarginal}, we illustrate the separation as a function of the turning point $\xi_0$ and the energy as a function of the separation.

\subsubsection*{The case of the dipole deformation II}

We again consider a configuration of a $D3$ brane with world-volume coordinates $(t, \, w, \, \phi, \, \phi_3)$, now embedded in the geometry of eq.~\eqref{metricTsT3w3}. As before, we assume that $w$ takes values in the interval $[-\nicefrac{d}{2}, \nicefrac{d}{2}]$ and that the $r$-direction has a non-trivial profile $r = r(w)$, subject to the boundary conditions \eqref{StringBCsMarginal}. The remaining coordinates are kept fixed, with $\th$ set to $\frac{\pi}{2}$ for consistency. This choice implies $\zeta = 1$, while the function $W$ becomes a non-trivial function of $r$, depending explicitly on the deformation parameter.

The induced metric on the $D3$ brane reads
\begin{equation}
    \begin{aligned}
  ds^2 & = - \frac{r^2}{L^2} dt^2
  + \left( \frac{r^2}{L^2 W^2}
  + \frac{r'^2}{r^2 F \lambda^6} \right) dw^2
  + r^2 F \, d\phi^2
  + \frac{L^2 \lambda^6}{W^2} \Big( d\phi_3 + \frac{A_3}{L} \Big)^2 \, .
 \end{aligned}
\end{equation}
Unlike the previous three cases, the function $W$ now depends on $r$. As a result, the dilaton and the NS two-form are non-trivial and contribute to the DBI action of the brane. The pull-back of the two-form onto the $D3$ brane is
\begin{equation}
    B_2 = \frac{\sqrt{W^2 - 1}}{W} \lambda^3 \, r \, dw \wedge \left( d\phi_3 + \frac{A_3}{L} \right) \, .
\end{equation}
The Wess--Zumino part of the action is trivial and, therefore, the dynamics of the brane is governed by the action \eqref{tHooftD3action}, where again $\cF$ and $\cG$ are given by eq. \eqref{FGtHooftMarginal}.

We see that, although the induced metric on the $D3$ brane differs from the previous cases, the non-trivial pull-back of the NS two-form and the dilaton conspire in such a way that the DBI action remains unchanged. In other words, the dynamics of the $D3$ brane is described by the same action for the original background as well as for both the marginal and dipole deformations. As a result, we find that the separation and the energy of the monopole-anti-monopole pair are not affected by the deformation parameter $\gamma$. Their behaviour is therefore expected to be the same as in the undeformed background \eqref{solutionANO}, studied in \cite{Chatzis:2025hek} and illustrated in Fig.~\ref{fig:tHooftMarginal}.


\subsection{Entanglement Entropy}

We now proceed with the study of holographic entanglement entropy (EE), following the prescription of \cite{Ryu:2006bv,Ryu:2006ef}. In particular, we consider the entangling region to be a strip of width $d$. The EE between the strip and its complementary region is obtained by determining a constant-time surface $\Sigma_8$ of minimal area whose boundary coincides with that of the strip. The action to be minimised is
\begin{equation}
\label{EEformula}
    S_{\text{EE}} = \frac{1}{4 G_{10}} \int_{\Sigma_8} d^8 \xi \, e^{- 2 \Phi} \sqrt{\det\!\left(g_{\Sigma_8}\right)} \, ,
\end{equation}
where $G_{10}$ is the ten-dimensional Newton's constant%
\footnote{This is related to $\alpha'$ and $g_s$ as $G_{10} = 8 \pi^6 \alpha'^4 g_s^2$.}, 
$g_{\Sigma_8}$ is the induced metric on $\Sigma_8$, and $\Phi$ is the dilaton. In the following, the eight-dimensional surface $\Sigma_8$ is taken to extend along the directions
\begin{equation}
\label{Sigma8}
    \xi^a = \left( w, z, \phi , \theta , \psi , \phi_1 , \phi_2 , \phi_3 \right) \, ,
\end{equation}
with a profile $r = r(z)$, where $z \in \left[ -\nicefrac{d}{2} , \nicefrac{d}{2} \right]$.

\begin{figure}[h ]
    \centering
    \includegraphics[width=0.7\textwidth]{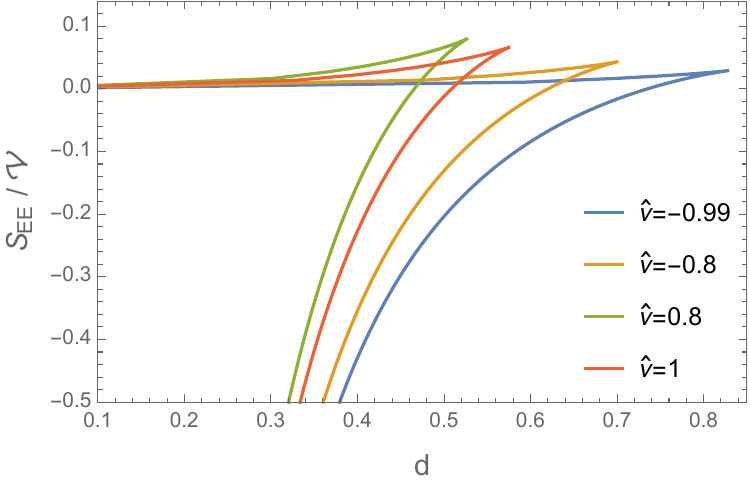}
    \caption{The entanglement entropy density as a function of the width, for different values of $\hat{\nu}$. For simplicity we set $L = \ell = 1$.}
    \label{fig:SEEmarginal}
\end{figure}

We expect the EE of the configuration \eqref{Sigma8} for the marginally deformed solutions \ref{solutionTsT121}, \ref{solutionTsT131}, as well as for the solutions dual to dipole deformations \ref{solutionTsT1w1} and \ref{solutionTsT3w3}, to behave exactly as in the case of the undeformed solution \ref{solutionANO}, which was studied in \cite{Chatzis:2025hek}. This follows from the fact that the integrand in \eqref{EEformula} is invariant under the TsT transformation, as can be readily verified by taking into account eqs.~\eqref{frameDeformationTsT121}, \eqref{frameDeformationTsT131}, \eqref{frameDeformationTsT1w1}, \eqref{frameDeformationTsT3w3}, and \eqref{dilatonTsT121}. In all cases, the EE functional \eqref{EEformula} reduces to the standard form
\begin{equation}
\label{SEEmarginal}
    S_{\text{EE}} = \cV \int_{- \nicefrac{d}{2}}^{\nicefrac{d}{2}} dz \, \sqrt{\cF^2 + \cG^2 \, r'^2} \, , \qquad
    \cV = \frac{\pi^3 L^4 L_w L_{\phi}}{4 G_{10}} \, ,
\end{equation}
where $\cF$ and $\cG$ are given in \eqref{FGtHooftMarginal}. The width $d$ can be expressed as a function of the turning point $\xi_0$, as in eq.~\eqref{qqbSeparation}. The EE in eq.~\eqref{SEEmarginal} is divergent and therefore must be regularised. The regularised EE density as a function of the width $d$ is shown in Fig.~\ref{fig:SEEmarginal}. The figure shows that the EE undergoes a phase transition, typical of confining theories \cite{Klebanov:2007ws,Kol:2014nqa,Georgiou:2015pia,Chatzis:2025hek}, where beyond a critical length the system exhibits a preference for a phase of vanishing EE. The two phases correspond to two disconnected Ryu--Takayanagi surfaces with boundary conditions $r\big( \pm \nicefrac{d}{2} \big) = \infty$.

The presence of a phase transition in the entanglement entropy is not unique to confining geometries. As seen in \cite{Jokela:2020wgs}, a phase transition in the entanglement entropy does not necessarily imply confinement. A TsT with the T-dual performed along the confining direction, $\phi$, will give a non-confining non-commutative geometry as the coefficient of $\mathrm{d}\phi^2$ will be deformed and the $S^1$ isometry is broken. However, this non-confining geometry will also have a phase transition in the entanglement entropy, it is the TsT which introduces a non-commutativity parameter \cite{Seiberg:1999vs},\cite{Lunin:2005jy} which introduces a length parameter into the metric. This causes the holographic EE to saturate because of a finite correlation length in the system. This modifies the UV in the dual field theory and the entanglement entropy exhibits a cross-over from the area law to a volume law. See \cite{Bea:2017iqt} for a similar study for non-commutative versions of the ABJM theory, where the TsT deformation generates UV modifications and a corresponding phase transition in the EE despite the absence of confinement.


\subsection{Holographic Central Charge Flow}

The last observable we would like to analyse is the flow of the holographic central charge. This is given in terms of a monotonic function of the holographic coordinate $r$, which captures the central charge in the UV and provides an effective account of the degrees of freedom along the RG flow. The flow of the central charge across dimensions can be determined holographically following the prescription of \cite{Bea:2015fja,Merrikin:2022yho}, which is summarised below.

We are interested in spacetimes supported by a dilaton, where both the metric and the dilaton take the form
\begin{equation}
\label{spacetimesForCC}
    \begin{aligned}
        ds^2 & = - \a_0 \, dt^2 + \sum\limits_{n = 1}^d \a_n \, dx^2_n + \left( \prod_{n = 1}^d \a_n \right)^{\frac{1}{d}} \beta \, dr^2 + g_{ij} \left( dy^i - A^i \right) \left( dy^j - A^j \right) \, , \\[5pt]
        \Phi & = \Phi(r , y^i) \, .
    \end{aligned}
\end{equation}
In the above expression $x_n \, (n = 1 , \ldots , d)$ are the spatial directions of the dual QFT and $y^i$ represent the coordinates of the $(8 - d)$-dimensional internal manifold with metric $g_{ij}$. We also allow for fibrations represented by the one-forms $A^i$. The holographic RG flow of the central charge is provided by the formula 
\begin{equation}
\label{cflowFormula}
    c_{\text{flow}} = d^d \frac{\beta^{\nicefrac{d}{2}} \, H^{\nicefrac{(2 d + 1)}{2}}}{G_{10} \, H'^d} \, , \qquad
    H := \left( \int d^dx \, d^{8-d}y \, e^{- 2 \Phi} \sqrt{\det(\hat{g}_{\S_8})} \right)^2 \, , 
\end{equation}
where the prime stands for derivation with respect to $r$, and $\hat{g}_{\S_8}$ corresponds to the metric of the subspace spanned by $(x_n , y^i)$ in eq. \eqref{spacetimesForCC}. In the following we study the $c_{\text{flow}}$ function for each of the TsT geometries presented in Sec. \ref{SupergravityBackgrounds}.

\subsubsection*{The case of the marginal deformations I \& II}

Starting with the marginal deformations of Secs.~\ref{solutionTsT121} and \ref{solutionTsT131}, we expect to recover the same results as in the undeformed background of Sec.~\ref{solutionANO}, which was studied in \cite{Chatzis:2025dnu,Chatzis:2025hek}. This is because the function $H$ in Eq.~\eqref{cflowFormula} is invariant under the TsT transformations, which in turn leave the $c_{\text{flow}}$ function itself invariant.

\begin{figure}[h!]
    \centering
\includegraphics[width=0.7\textwidth]{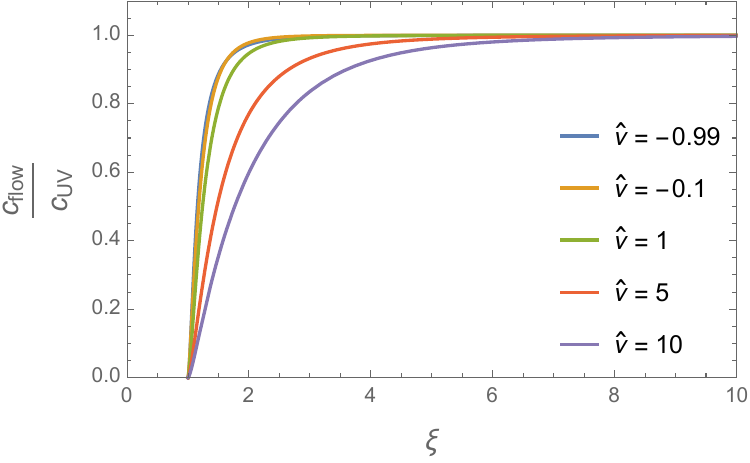}
    \caption{The flow of the central charge for different values of $\hat{\nu}$.}
    \label{fig:CCmarginal}
\end{figure}

For both marginally deformed backgrounds we have $d = 3$, while the functions $\a_0 , \ldots , \a_3$ and $\b$ read
\begin{equation}
    \a_0 = \ldots = \a_2 = \frac{r^2 \, \zeta}{L^2} \, , \qquad
    \a_3 = r^2 F \zeta \, , \qquad
    \b = \frac{L^{\nicefrac{4}{3}}}{r^4 F^{\nicefrac{4}{3}} \l^6} \, .
\end{equation}
For the function $H$ of eq. \eqref{cflowFormula} we find
\begin{equation}
    H = \left( L_w L_z L_{\phi} \right)^2 \left( \pi \, L \, r \, \l \right)^6 F \, ,
\end{equation}
where we notice that it depends only on $r$ as the dependence on $\th$ cancels out. For the flow of the central charge we obtain
\begin{equation}
    c_{\text{flow}} = 27 \, \cN \, \frac{\sqrt{\left( \hat{\nu} + \xi^2 \right) \left( \xi^6 - 1 + \hat{\nu} \left( \xi^4 - 1 \right) \right)^3}}{\xi^4 \left( 2 \hat{\nu} + 3 \xi^2 \right)^3} \, , \qquad
    \cN := \frac{\pi^3 L^8 L_w L_z L_{\phi}}{8 \, G_{10}} \, ,
\end{equation}
where for convenience we expressed the result in terms of the coordinate $\xi = \frac{r}{r_\star}$. Therefore the central charge interpolates between the following values
\begin{equation}
    c_{\text{UV}} = \lim\limits_{\xi \to \infty} c_{\text{flow}} = \cN \, , \qquad
    c_{\text{IR}} = \left. c_{\text{flow}} \right|_{\xi = 1} = 0 \, .
\end{equation}
The vanishing of the function in the IR indicates that the theory is gapped. The behaviour of the central charge flow is monotonically increasing towards the UV. This is illustrated for different values of the parameter $\hat{\nu}$ in Fig. \ref{fig:CCmarginal}.

We see that the observables have the same universality properties as in \cite{Chatzis:2025dnu,Chatzis:2025hek}.

\subsubsection*{The case of the dipole deformations I \& II}

Applying the formulas \eqref{spacetimesForCC} and \eqref{cflowFormula} to the solutions of Sec.~\ref{solutionTsT1w1} and Sec.~\ref{solutionTsT3w3}, one obtains a $c$-function that depends explicitly on some of the internal coordinates. This indicates that the prescription for computing the holographic central charge flow is not directly applicable to spacetimes of this type and therefore requires generalisation.

\section{Conclusions}

In this paper, we constructed four type-IIB supergravity solutions by applying TsT transformations to the uplift of the five-dimensional soliton solution of \cite{Anabalon:2024che}. Two of the resulting backgrounds are holographic duals of marginal deformations of the QFT associated with \cite{Anabalon:2024che}, while the remaining two are holographic duals of dipole deformations of the same QFT. In order to get insights into the QFT side we studied a number of observables including Page charges, Wilson loops, 't Hooft loops, entanglement entropy and holographic central charge flow.

Regarding the Page charges, we found that in the case of the marginal deformations the TsT transformation generates an additional charge associated with the presence of $D5$ branes, beyond the $D3$ brane charge inherited from the seed solution. The $D5$ charge is proportional to the $D3$ charge, with the deformation parameter $\gamma$ serving as the proportionality constant. Charge quantisation then suggests that $\gamma$ must take rational values. Unlike the marginally deformed examples, in the dipole-deformed cases the TsT transformation does not generate additional brane charges. Consequently, these solutions carry the same charges as the seed background.

The Wilson loop computation is performed for the same string configuration as in \cite{Chatzis:2025dnu,Chatzis:2025hek}, allowing for direct comparison. We find that, in the case of the marginal deformations, the corresponding Wilson loop behaves exactly as in the original background. This is no longer true for the dipole deformations, where the Wilson loop exhibits an explicit dependence on the deformation parameter $\gamma$. In these cases, the quark-anti-quark energy develops a \emph{wedge} behaviour at small separations, which may be related to the absence of conformality in the UV when the deformation is present. Towards the IR, the energy grows linearly, indicating a confining behaviour. Moreover, for the example corresponding to the geometry of Sec. \ref{solutionTsT1w1}, we observed that for sufficiently large values of the parameters $\hat{\nu}$ and $\gamma$ the energy of the pair exhibits a \emph{triangular} behaviour, suggesting the presence of a first-order phase transition.

Continuing with the 't Hooft loop, we analysed the dynamics of a $D3$ brane with world-volume coordinates $(t, \, w, \, \phi, \, \phi_3)$, following \cite{Chatzis:2025hek} for the seed solution. We find that the energy of the magnetic monopole-anti-monopole pair is unaffected by the TsT transformation in all TsT-generated backgrounds. 

An analogous conclusion holds for the entanglement entropy of a strip extending along the $z$ direction, with $z \in [ - \nicefrac{d}{2} , \nicefrac{d}{2} ]$. However, we expect this not to remain true when the strip extends along the $w$ direction. In that case, the entanglement entropy should be sensitive to the deformation, at least for the geometries corresponding to dipole deformations. Such a computation is technically more involved, as it requires non-trivial integrations over internal coordinates in the evaluation of the entropy density. We therefore leave this analysis for future work.

The final observable we considered is the holographic central charge flow, following the prescription of \cite{Bea:2015fja,Merrikin:2022yho}. In the case of the marginal deformations, we found that this quantity is invariant under the TsT transformation and therefore reproduces the behaviour of the undeformed background. On the other hand, for the dipole deformations we concluded that the existing prescription for computing the central charge flow requires generalisation in order to be consistently applied to these geometries \footnote{We are aware that work on generalising the formula for the holographic flow central charge is currently in progress.}.

The seed geometry \eqref{metricANO} exhibits six $U(1)$ isometries in total, three associated with external coordinates and the remaining three with internal ones. This provides considerable freedom in constructing TsT-transformed solutions. In the present work we have not exhausted all such possibilities, and several interesting cases, such as TsT deformations dual to non-commutative QFTs or transformations involving the confining direction $\phi$, are left for future investigation. Additional directions that we plan to explore include the stability analysis of Wilson loops and the study of Krylov complexity for the TsT-generated backgrounds.

A construction based on a type-IIB background with features similar to those of Sec. \ref{solutionANO} was considered in \cite{Gursoy:2005cn}. In that work, the authors applied a TsT transformation to a solution with topology $\mathbb{R}^{1,3} \times S^2 \times S^3 \times \mathbb{R}$ preserving four supercharges \cite{Nunez:2003cf}, where the field-theory directions extend along $\mathbb{R}^{1,3} \times S^2$. The TsT transformation was performed along two $U(1)$ directions embedded in the spheres $S^2$ and $S^3$, generating a dipole deformation in the dual QFT. Evidence was presented that the deformation affects only the Kaluza-Klein sector, namely the sector charged under the two $U(1)$ symmetries. It would be interesting to investigate whether a similar mechanism operates for TsT deformations of the solution in Sec. \ref{solutionANO}, particularly when the transformation involves the compactified field-theory direction $\phi$ and one internal $U(1)$ direction.


\section{Acknowledgements}
We are grateful to Carlos Nu\~nez for carefully reading the draft and providing useful comments and insights. We also thank Dimitrios Chatzis, Prem Kumar, Dimitrios Zoakos, Niko Jokela and Federico Castellani for discussions. M.H thanks Durham University YTF25 for hospitality and for the chance to speak about this work. M.H. has been supported by the STFC consolidated grant ST/Y509644/1. The research of G.I. is supported by the Einstein Stiftung Berlin via the Einstein International Postdoctoral Fellowship program ``Generalised dualities and their holographic applications to condensed matter physics'' (project number IPF- 2020-604).



\bibliography{main.bib}
\bibliographystyle{utphys}

\end{document}